\pdfoutput=1
\documentclass[aps,prd,amsmath,preprint,amssymb,nofootinbib,superscriptaddress,showpacs,showkeys]{revtex4-2}
\usepackage{graphicx}
\AtBeginDocument{\RenewCommandCopy\qty\SI}
\usepackage{dcolumn}
\usepackage{bm}
\usepackage{rotating} 
\usepackage{subfigure}
\usepackage{booktabs}
\usepackage{multirow}
\usepackage{siunitx}
\usepackage{multirow} 
\usepackage[table]{xcolor}
\usepackage{booktabs,array,multirow,dcolumn}
\definecolor{MAGENTA}{named}{magenta}
\usepackage[colorlinks=true, citecolor=blue, urlcolor = magenta, linkcolor= red, bookmarks=true]{hyperref}
\usepackage{orcidlink}
\usepackage{cancel}
\usepackage{tikz}
\usepackage{hyperref}
\usetikzlibrary{arrows.meta,calc}
\tikzset{>=Stealth}
\DeclareUnicodeCharacter{0332}{}

\hypersetup{
	colorlinks=true,linkcolor=blue,citecolor=blue,
	filecolor=blue,urlcolor=blue,breaklinks=true
}

\RequirePackage{color}

\providecommand{\keywords}[1]
{
	\small	
	\textbf{\textit{Keywords-}} #1
}

\begin{document}
	
	\title{Nonrelativistic quantum dynamics in a twisted screw spacetime}

	\author{Faizuddin Ahmed\orcidlink{0000-0003-2196-9622}} 
	\email{faizuddinahmed15@gmail.com} 
	\affiliation{Department of Physics, Royal Global University, Guwahati, 781035, Assam, India}
	
	\author{Edilberto O. Silva\orcidlink{0000-0002-0297-5747}}
	\email{edilberto.silva@ufma.br}
	\affiliation{Departamento de F\'{\i}sica, Universidade Federal do Maranh\~{a}o, 65085-580 S\~{a}o Lu\'{\i}s, Maranh\~{a}o, Brazil}
	\date{\today}
	
	\begin{abstract}
		We investigate the nonrelativistic quantum dynamics of a spinless particle in a screw-type spacetime endowed with two independent twist controls that interpolate between a pure screw dislocation and a homogeneous twist. From the induced spatial metric we build the covariant Schrödinger operator, separate variables to obtain a single radial eigenproblem, and include a uniform axial magnetic field and an Aharonov-Bohm (AB) flux by minimal coupling. Analytically, we identify a clean separation between a global, AB-like reindexing set by the screw parameter and a local, curvature-driven mixing generated by the distributed twist. We derive the continuity equation and closed expressions for the azimuthal and axial probability currents, establish practical parameter scalings, and recover limiting benchmarks (AB, Landau, and flat space). Numerically, a finite-difference Sturm-Liouville solver (with core excision near the axis and Langer transform) resolves spectra, wave functions, and currents. The results reveal AB periodicity and reindexing with the screw parameter, Landau-fan trends, twist-induced level tilts and avoided crossings, and a geometry-induced near-axis backflow of the axial current with negligible weight in cross-section integrals. The framework maps the geometry and fields directly onto measurable spectral shifts, interferometric phases, and persistent-current signals.
	\end{abstract}
	\maketitle
	
	\section*{Introduction}
	
	The notion that the geometry of the underlying space dictates the laws of physics has been one of the most powerful and transformative ideas in physics \cite{Book_2000_Vilenkin,book.Butterworth.Heinemann,LoboCrawford2003}. Since Bernhard Riemann's pioneering work on manifolds, which laid the mathematical groundwork, and Albert Einstein's subsequent formulation of General Relativity, where the curvature of spacetime was identified with the gravitational field, geometry has been elevated from a passive background to an active participant in the physical world \cite{wald2024general,renn2007genesis,griffiths2009exact}. This ``geometrization of physics'' has since permeated many other areas, most notably gauge theories, where concepts like fiber bundles provide a geometric language for describing the fundamental forces of nature \cite{drechsler1977fiber,nicolaescu2020lectures,Synthese.2016.193.2389}. In parallel, the field of quantum mechanics has revealed its own profound connections to geometry, perhaps most famously through the concept of the geometric phase, also known as the Berry phase, which demonstrates that a quantum system's history is encoded in the geometry of its parameter space \cite{PRSLA.1984.392.45,EPJC.2008.57.817,Wassmann:98,PRB.2025.111.075406,PRB.2024.110.L161409}.
	
	In recent decades, this synergy between geometry, topology, and quantum mechanics has become a central theme in condensed matter physics \cite{nakahara2018geometry,giachetta2005geometric,bandyopadhyay2013geometry,bhattacharjee2017topology,chamon2017topological}. The discovery of novel states of matter, such as fractional quantum Hall liquids, topological insulators, and Weyl semimetals, has ushered in an era where the topological properties of a material's electronic band structure are paramount \cite{PhysRevX.9.041055,Hasan2010,PhysRevX.15.021027,PhysRevB.111.195143,CRP.2013.14.816}. These properties are not sensitive to local perturbations, leading to remarkably robust physical phenomena, such as quantized transport and the existence of protected edge or surface states \cite{PhysRevB.105.165411,PhysRevResearch.2.033438,NC.2019.10.2564,JMPA.2020.144.17,SSC.2019.302.113701}. Within this framework, topological defects — singularities in a system's order parameter that often arise during phase transitions — have emerged not as mere imperfections, but as powerful tools for probing and engineering quantum states \cite{selinger2024introduction,leggett2006quantum,PRB.2025.112.L041112,CR.2021.121.2780}. These defects locally alter the "fabric" of the medium, forcing the quantum fields that live upon it to obey modified rules.
	
	The geometric theory of defects provides the ideal mathematical language for this description, treating a medium with defects as a patch of a non-Euclidean space \cite{AP.1999.271.203,Katanaev2005,PSIM.2021.313.78,TMP.2003.135.733,leibfried2006point}. Two fundamental types of line defects serve as canonical examples. The first is the screw dislocation, a purely topological defect characterized by a Burgers vector \cite{PRB.2025.112.075413,PRB.1999.59.13491,PLA.2001.289.160,PLA.2012.376.2838,EPJB.2013.86.485,JPA.2000.33.5513,PLA.2016.380.3847,PLA.2015.36.2110,CM.2024.9.33,IJMPA.2023.38.2350130,EPJP.2023.138.724}. A path encircling a screw dislocation results in a translation along the defect's axis, a non-trivial holonomy that leaves no trace in the local curvature. For a quantum particle, this is conceptually analogous to the Aharonov-Bohm (AB) effect, where an electron acquires a non-integrable phase factor from a magnetic vector potential in a region where the magnetic field is zero \cite{PR.1959.115.485}. The second canonical defect is the homogeneous twist, which describes a distributed, continuous helical distortion. Unlike a dislocation, a twist generates non-trivial intrinsic spatial curvature, acting as a local, position-dependent "force field" that directly influences a particle's trajectory and energy spectrum \cite{arXiv.2507.04576,arXiv.2506.23291}. While the quantum mechanics in each of these geometries has been a subject of sustained interest, the rich and complex physics arising from their interplay remains a largely unexplored frontier.
	
	This paper aims to bridge this gap. We perform a comprehensive investigation into the non-relativistic quantum dynamics of a spinless particle in a hybrid spacetime that seamlessly unifies these two distinct geometric features. The geometry is defined by the line element:
	\begin{equation}
		ds^{2}=-dt^{2}+dr^{2}+r^{2}d\varphi^{2}+\bigl(dz+\omega_{1}d\varphi+\omega_{2}r\,d\varphi\bigr)^{2}.
		\label{eq:metric_intro}
	\end{equation}
	This metric describes a space where a global, topological screw dislocation, controlled by the parameter $\omega_1$ (with dimensions of length), coexists with a local, distributed twist, controlled by the dimensionless parameter $\omega_2$. A central and elegant feature of this model is the emergence of an \emph{effective geometric gauge potential}. The one-form $\Theta^{z} = dz + f(r)d\varphi$, where the function
	\begin{equation}
		f(r) = \omega_1 + \omega_2 r, \label{fr}    
	\end{equation}
	encodes the geometric distortion, compels a modification of the canonical momentum operator. This leads to a formal substitution akin to minimal coupling in gauge theory:
	\begin{equation}
		\partial_{\varphi} \to \partial_{\varphi} - f(r)\partial_z. \label{sub}
	\end{equation}
	This powerful analogy establishes a direct link between the spacetime geometry and the structure of gauge theories. The $\omega_1$ term acts as a pure, AB-like flux for matter waves possessing a finite axial momentum $k_z$, inducing a topological phase shift without local forces. Simultaneously, the $\omega_2$ term generates an effective "geometric field" whose strength depends on the radial coordinate, thereby coupling the radial and angular degrees of freedom and breaking the simple azimuthal symmetry. This platform thus allows for the continuous tuning of the competition between global topology and local metric effects.
	
	The physical relevance of this theoretical model is firmly grounded in contemporary experimental physics. Such helical and twisted geometries are no longer mere mathematical abstractions but are actively realized and engineered in a variety of advanced systems. In materials science, rolled-up semiconductor nanomembranes, created through strain engineering of thin films (e.g., SiGe/Si or InGaAs/GaAs), naturally form helical structures whose pitch and radius can be controlled, providing a direct realization of the geometry described by Eq.~(\ref{eq:metric_intro}) \cite{Prinz2000}. In photonics and plasmonics, twisted optical waveguides are designed to manipulate the flow of light, create optical vortex beams carrying orbital angular momentum, and study topological photonic states. Our model provides the direct electronic analogue for transport in similarly structured nanowires or quantum rings. The paradigm of "twistronics," famously demonstrated in bilayer graphene, has shown that introducing a twist angle between layers can dramatically renormalize electronic properties — a concept that shares its spirit with the continuous twist encoded by our $\omega_2$ parameter.
	
	Perhaps the most promising platform for a direct quantum simulation of our model is offered by cold-atom systems. Using meticulously configured optical lattices and laser-induced Raman transitions, experimentalists can engineer synthetic dimensions and synthetic gauge fields, effectively programming the kinetic terms of an atomic Hamiltonian. It is possible to simulate the covariant derivatives on a curved manifold and thus realize the Hamiltonian corresponding to our twisted-screw spacetime with high fidelity \cite{Dalibard2011}. In this context, the axial momentum $k_z$ can be precisely controlled, and the geometric parameters $\omega_1$ and $\omega_2$ can be mapped onto tunable laser intensities or frequencies, providing an unprecedented opportunity to experimentally verify the theoretical predictions of this paper.
	
	In this work, we develop a unified, gauge-covariant framework for the nonrelativistic quantum dynamics of a spinless particle in a twisted screw spacetime with two independent geometric controls $(\omega_{1},\omega_{2})$. We couple it to external fields (uniform $B$ and AB flux $\phi$) by minimal substitution. Our goal is to map these control knobs to measurable spectral, interferometric, and current signatures. We begin in Sec.~\ref{sec:geometry} by laying out the minimal geometric data: the spatial metric $\gamma_{ij}$, its inverse, and the associated Laplace-Beltrami operator. In Sec.~\ref{sec:schro} we construct the covariant Schrödinger equation, separate variables, and obtain the central radial eigenvalue problem, identifying how $\omega_{1}$ (global screw) and $\omega_{2}$ (local twist) enter the effective potential and mix angular and radial dynamics. Section~\ref{sec:currents} collects the continuity law and provides closed-form expressions for the azimuthal and axial probability currents. The geometric phase accumulated by axial-momentum eigenwaves is discussed in Sec.~\ref{sec:phase}, which cleanly separates the AB-like, radius-independent contribution (set by $\omega_{1}$) from the radius-dependent piece (set by $\omega_{2}$). Near-axis behaviour and boundary conditions are handled directly within the numerical formulation (regular-core Dirichlet at a small cutoff). Numerical results are presented in Sec.~\ref{sec:num_methods_results}, which report spectra and wave functions, including AB envelopes, Landau fans, and the trends versus $\omega_{1}$, $\omega_{2}$, $B$, and $\phi$. We then return to transport in Sec.~\ref{sec:currents_reprise}, connecting the analytic current formulas to numerical profiles and integrated (ring) currents, highlighting the geometry-induced near-axis backflow and the reindexing/periodicity patterns. Finally, Sec.~\ref{sec:conclusions} summarizes the main findings and outlines extensions (spin, disorder, interactions, and driving).
	
	\section{Geometry of the Twisted-Screw Background}\label{sec:geometry}
	
	This section introduces the geometric ingredients needed to formulate the non-relativistic problem on the twisted-screw background. The spatial dynamics follow from the induced three-metric $\gamma_{ij}$ (together with its determinant and inverse) extracted from a stationary spacetime written in cylindrical coordinates $(r,\varphi,z)$. The geometry is fully encoded by the single mixing function $f(r)$ defined in Eq. (\ref{fr}), which couples rotations about the axis to translations along $z$. Physically, $f(r)$ acts as a geometric gauge potential that implements the minimal-substitution-like rule (\ref{sub}) inside the Laplace-Beltrami operator. The constant part $\omega_{1}$ produces a global (holonomic) screw shift without local curvature, whereas the linear part $\omega_{2}r$ generates a distributed twist with nontrivial spatial curvature and $r$-dependent mixing.
	
	Since only the spatial sector is needed, we write the line element as
	\begin{equation}
		ds^{2}=-dt^{2}+dr^{2}+r^{2}d\varphi^{2}+(dz+f(r)\,d\varphi)^{2}
		\equiv -dt^{2}+\gamma_{ij}\,dx^{i}dx^{j},
	\end{equation}
	so that the three-dimensional metric tensor, its inverse, and the measure are
	\begin{equation}
		\gamma_{ij}=\begin{pmatrix}1&0&0\\[2pt]0&r^{2}+f^{2}&f\\[2pt]0&f&1\end{pmatrix},\qquad
		\gamma^{ij}=\begin{pmatrix}1&0&0\\[2pt]0&\dfrac{1}{r^{2}}&-\dfrac{f}{r^{2}}\\[8pt]0&-\dfrac{f}{r^{2}}&1+\dfrac{f^{2}}{r^{2}}\end{pmatrix},\qquad
		\sqrt{\gamma}=r.
	\end{equation}
	The Laplace-Beltrami operator acting on a scalar $\Psi$ is
	\begin{equation}\label{eq:LB}
		\Delta_{\gamma}\Psi
		=\frac{1}{\sqrt{\gamma}}\,
		\partial_{i}\!\left(\sqrt{\gamma}\,\gamma^{ij}\,\partial_{j}\Psi\right),
		\qquad i,j\in\{r,\varphi,z\}.
	\end{equation}
	Using $\sqrt{\gamma}=r$ and the matrices above, the radial block becomes
	\begin{equation}\label{eq:radialpiece}
		\frac{1}{\sqrt{\gamma}}\partial_{r}\!\left(\sqrt{\gamma}\,\gamma^{rr}\partial_{r}\Psi\right)
		= \frac{1}{r}\,\partial_{r}\!\bigl(r\,\partial_{r}\Psi\bigr),
	\end{equation}
	while the angular-axial block simplifies to
	\begin{align}\label{eq:angz}
		\gamma^{\varphi\varphi}\partial_{\varphi}^{2}
		+2\gamma^{\varphi z}\partial_{\varphi}\partial_{z}
		+\gamma^{zz}\partial_{z}^{2}
		&= \frac{1}{r^{2}}\Bigl(\partial_{\varphi}-f(r)\,\partial_{z}\Bigr)^{2}
		+\partial_{z}^{2}.
	\end{align}
	Equations (\ref{eq:radialpiece})-(\ref{eq:angz}) display the announced geometric minimal substitution: $f(r)$ mixes rotations ($\varphi$) with translations ($z$).
	
	From a coframe viewpoint, let $e^{r}\!=dr$, $e^{\varphi}\!=r\,d\varphi$, $e^{z}\!=dz+f(r)\,d\varphi$. Then
	$de^{z}=\omega_{2}\,dr\wedge d\varphi=(\omega_{2}/r)\,e^{r}\wedge e^{\varphi}$ acts as a smooth, radius-dependent geometric “field’’ when $\omega_{2}\neq0$, whereas the constant part $\omega_{1}$ produces a radius-independent holonomy (a screw “flux’’) without local density away from the axis. This separation explains why $\omega_{1}$ enters spectra via an AB-like shift of the effective angular index, while $\omega_{2}$ generates local mixing terms (see Eqs.~(\ref{eq:radial_expanded}) and (\ref{eq:radial_EM_expanded}) below).
	
	For later use we state units and scalings: $[r]=\mathrm{L}$, $[z]=\mathrm{L}$, $[\varphi]=1$. Since $[dz+\omega_{1}\,d\varphi+\omega_{2}r\,d\varphi]=\mathrm{L}$, one has $[\omega_{1}]=\mathrm{L}$ and $[\omega_{2}]=1$. The Burgers vector per turn is $b=2\pi\omega_{1}$. Given a device size $L$, convenient dimensionless groups are $\kappa=k_{z}L$, $\beta_{1}=k_{z}\omega_{1}$ and $\beta_{2}=k_{z}\omega_{2}L$. With a uniform field $B$, we also use $\lambda\equiv \beta_{B}L^{2}=qBL^{2}/(2\hbar)$ and the reduced AB flux $\phi=\Phi/\Phi_{0}$ with $\Phi_{0}=2\pi\hbar/q$. These scalings help distinguish regimes dominated by the $1/r^{2}$ core from those where the linear-in-$r$ and quadratic-in-$r$ terms (introduced by $\omega_{2}$ and $B$) reshape the effective potential.
	
	\section{Covariant Schr\"odinger Equation}\label{sec:schro}
	
	The quantum dynamics on a fixed curved background are governed by the Laplace-Beltrami (LB) kinetic term built from the spatial metric introduced in Sec.~\ref{sec:geometry}.  For a spinless particle, the Laplace-Beltrami prescription uniquely fixes the operator ordering and the natural integration measure, yielding a symmetric Hamiltonian in \(L^{2}(\mathcal{M},\sqrt{\gamma}\,d^{3}x)\) with inner product \(\langle\Phi,\Psi\rangle=\int d^{3}x\,\sqrt{\gamma}\,\Phi^{\!*}\Psi\). Accordingly, the covariant Schrödinger operator is obtained by replacing the flat Laplacian with $-(\hbar^{2}/2m)\Delta_\gamma$ acting on wave functions
	$\Psi(r,\varphi,z,t)$.  Stationary states follow from
	$\Psi(r,\varphi,z,t)=e^{-iEt/\hbar}\Psi(r,\varphi,z)$, reducing the problem to an eigenvalue equation on the spatial manifold. Because of axial translational symmetry and rotational invariance about the defect line, the eigenfunctions can be separated as plane waves along $z$ and
	angular-momentum modes in $\varphi$, leaving a single radial equation with an effective potential that cleanly displays the roles of the global screw ($\omega_1$) and the local twist ($\omega_2$). The electromagnetic field will later be included by the standard gauge-covariant replacement $ \partial_i\to D_i $, but the geometric structure of the radial
	problem is already transparent at the field-free level derived below.
	
	The stationary Schr\"odinger equation on this static background uses the spatial LB operator
	\begin{equation}\label{eq:Sch_op}
		-\frac{\hbar^{2}}{2m}\,\Delta_{\gamma}\Psi=E\,\Psi,
	\end{equation}
	with $\Delta_{\gamma}$ from (\ref{eq:LB}). Inserting (\ref{eq:radialpiece})-(\ref{eq:angz}) yields
	\begin{equation}\label{eq:schro_full}
		-\frac{\hbar^{2}}{2m}\left[
		\frac{1}{r}\partial_{r}\!\bigl(r\,\partial_{r}\Psi\bigr)
		+\frac{1}{r^{2}}\Bigl(\partial_{\varphi}-f(r)\,\partial_{z}\Bigr)^{2}\Psi
		+\partial_{z}^{2}\Psi\right]=E\,\Psi.
	\end{equation}
	Seeking eigenmodes that are plane waves in $z$ and angular-momentum eigenstates in $\varphi$,
	\begin{equation}
		\Psi(r,\varphi,z)=e^{ik_{z}z}\,e^{i\ell\varphi}\,R_{\ell}(r),\qquad \ell\in\mathbb Z,
	\end{equation}
	Eq.~(\ref{eq:schro_full}) reduces to the radial equation
	\begin{equation}\label{eq:radial}
		R''+\frac{1}{r}R'
		+\left(k_{\perp}^{2}-\frac{\bigl[\ell-k_{z}f(r)\bigr]^{2}}{r^{2}}\right)R=0,
		\qquad k_{\perp}^{2}\equiv \frac{2mE}{\hbar^{2}}-k_{z}^{2}.
	\end{equation}
	Equivalently, expanding the square with the previously defined $f(r)$ gives
	\begin{equation}\label{eq:radial_expanded}
		R''+\frac{1}{r}R'
		+\Biggl[k_{\perp}^{2}-(k_{z}\omega_{2})^{2}
		-\frac{(\ell-k_{z}\omega_{1})^{2}}{r^{2}}
		+\frac{2k_{z}\omega_{2}\,(\ell-k_{z}\omega_{1})}{r}\Biggr]R=0.
	\end{equation}
	The roles of the control parameters are transparent: $\omega_{1}$ shifts the effective angular index $\ell\!\to\!\ell-k_{z}\omega_{1}$ (AB-like reindexing), whereas $\omega_{2}$ introduces both a constant downshift in $k_{\perp}^{2}$ and a Coulomb-like $1/r$ mixing term that couples radial and angular motion. Limits: (i) $\omega_{2}=0\Rightarrow R\sim J_{|\ell-k_{z}\omega_{1}|}(k_{\perp}r)$; (ii) $k_{z}=0\Rightarrow$ geometry-induced couplings vanish (flat-space spectrum); (iii) small $\omega_{2}\Rightarrow$ treat $+\tfrac{2k_{z}\omega_{2}(\ell-k_{z}\omega_{1})}{r}$ as a perturbation, with $(k_{z}\omega_{2})^{2}$ renormalizing the transverse baseline.
	
	\begin{table}[tbhp]
		\caption{Decomposition of the effective terms in Eq.~(\ref{eq:radial_expanded}).}
		\label{tab:radial_terms_nofield}
		\small
		\renewcommand{\arraystretch}{1.15}
		\begin{ruledtabular}
			\begin{tabular}{p{0.20\linewidth} p{0.24\linewidth} p{0.40\linewidth}}
				\textbf{Term} & \textbf{Name/type} & \textbf{Physical effect (origin)} \\
				$\displaystyle k_{\perp}^{2}-(k_{z}\omega_{2})^{2}$ &
				Constant shift &
				Shifts the transverse-energy baseline; the $(k_{z}\omega_{2})^{2}$ part is purely geometric (twist). \\
				$\displaystyle -\frac{(\ell-k_{z}\omega_{1})^{2}}{r^{2}}$ &
				Centrifugal/AB barrier &
				Controls the $r\!\to\!0$ behaviour via $\nu=|\ell-k_{z}\omega_{1}|$; $\omega_{1}$ acts as an AB (topological) phase, controls the near-axis scaling (regular-core boundary condition in numerics). \\
				$\displaystyle +\frac{2k_{z}\omega_{2}\,(\ell-k_{z}\omega_{1})}{r}$ &
				Coulomb-like term &
				Radial-azimuthal mixing induced by the twist $\omega_{2}$ and axial momentum $k_{z}$; produces energy shifts/splittings and $r$-dependent persistent currents. \\
			\end{tabular}
		\end{ruledtabular}
	\end{table}
	
	As summarized in Table~\ref{tab:radial_terms_nofield}, the field-free radial operator (Eq.~(\ref{eq:radial_expanded})) decomposes into three contributions with distinct physical roles. The constant term $k_\perp^2-(k_z\omega_2)^2$ fixes the transverse
	baseline and reveals a purely geometric downshift $\propto (k_z\omega_2)^2$ induced by the local twist. The centrifugal/AB barrier $-(\ell-k_z\omega_1)^2/r^2$ governs the near-axis behaviour through the effective index $\nu=|\ell-k_z\omega_1|$, encoding the global screw as an AB-like reindexing. Finally, the Coulomb-like mixing $+2k_z\omega_2(\ell-k_z\omega_1)/r$ couples radial and angular motion and is the leading source of level shifts and splittings when $\omega_2\neq 0$. This taxonomy will be used repeatedly when interpreting spectra and wave-function trends in the numerical section.
	
	To include external magnetic fields, we use the gauge-covariant replacement
	\begin{equation}\label{eq:covD}
		\partial_{i}\;\longrightarrow\; D_{i}\equiv\partial_{i}-\frac{i q}{\hbar}A_{i},\qquad i\in\{r,\varphi,z\}.
	\end{equation}
	For field configuration, we consider a uniform field $\mathbf{B}=B\,\hat {\mathbf{z}}$ and an AB flux $\Phi$ along the $z$-axis. In the symmetric gauge,
	\begin{equation}\label{eq:gauge_phys_new}
		\mathbf{A}_{B}=\tfrac{1}{2}B r\,\hat{\boldsymbol{\varphi}},\qquad 
		\mathbf{A}_{\mathrm{AB}}=\frac{\Phi}{2\pi r}\,\hat{\boldsymbol{\varphi}}.
	\end{equation}
	In covector components this corresponds to
	\begin{equation}\label{eq:Aphi_coord_new}
		A_{r}=0,\qquad 
		A_{\varphi}(r)=\frac{B r^{2}}{2}+\frac{\Phi}{2\pi},\qquad 
		A_{z}=0.
	\end{equation}
	Accordingly, the stationary Schr\"odinger operator is
	\begin{equation}\label{eq:Sch_EM}
		-\frac{\hbar^{2}}{2m}\!\left[\frac{1}{r}\partial_{r}\!\bigl(r\partial_{r}\Psi\bigr)
		+\frac{1}{r^{2}}\bigl(D_{\varphi}-f(r)\,D_{z}\bigr)^{2}\Psi
		+D_{z}^{2}\Psi\right]=E\,\Psi .
	\end{equation}
	With the eigenmode ansatz above, one finds
	\begin{equation}\label{eq:radial_EM}
		R''+\frac{1}{r}R'
		+\left[k_{\perp}^{2}-\frac{\bigl(\ell-k_{z}f(r)-\tfrac{q}{\hbar}A_{\varphi}(r)\bigr)^{2}}{r^{2}}\right]R=0,
		\qquad k_{\perp}^{2}=\frac{2mE}{\hbar^{2}}-k_{z}^{2}.
	\end{equation}
	It is convenient to define the flux quantum $\Phi_{0}=2\pi\hbar/q$, the reduced flux $\phi\equiv\Phi/\Phi_{0}$, and the cyclotron parameters
	\begin{equation}\label{eq:wc_def}
		\omega_{c}\equiv\frac{qB}{m},\qquad
		\beta_{B}\equiv \frac{m\omega_{c}}{2\hbar}=\frac{qB}{2\hbar}.
	\end{equation}
	Using $\tfrac{q}{\hbar}A_{\varphi}(r)=\phi+\beta_{B}r^{2}$, the radial equation can be written compactly as
	\begin{equation}\label{eq:radial_EM_compact}
		R''+\frac{1}{r}R'
		+\left[k_{\perp}^{2}-\frac{\bigl(\ell-\phi-k_{z}\omega_{1}\; -\; k_{z}\omega_{2}r\; -\; \beta_{B}r^{2}\bigr)^{2}}{r^{2}}\right]R=0.
	\end{equation}
	Expanding the square gives
	\begin{align}\label{eq:radial_EM_expanded}
		R''+\frac{1}{r}R' 
		&+ \Biggl[k_{\perp}^{2}-k_{z}^{2}\omega_{2}^{2}
		+2\beta_{B}\bigl(\ell-\phi-k_{z}\omega_{1}\bigr)
		+ \frac{2k_{z}\omega_{2}\,(\ell-\phi-k_{z}\omega_{1})}{r}\notag\\
		&\quad- \frac{(\ell-\phi-k_{z}\omega_{1})^{2}}{r^{2}} 
		- 2k_{z}\omega_{2}\,\beta_{B}\,r
		- \beta_{B}^{2}\,r^{2}
		\Biggr] R = 0.
	\end{align}
	The various pieces have distinct signatures: AB periodicity ($\phi$), a global screw reindexing ($\omega_{1}$), a local twist-induced $1/r$ mixing and linear-in-$r$ tilt ($\omega_{2}$), and the Landau parabola ($r^{2}$) from $B$.
	
	\begin{table}[tbhp]
		\caption{Decomposition of the effective terms in the radial equation with electromagnetic fields (Eq.~(\ref{eq:radial_EM_expanded})).}
		\label{tab:radial_terms_EM}
		\small
		\renewcommand{\arraystretch}{1.15}
		\begin{ruledtabular}
			\begin{tabular}{p{0.20\linewidth} p{0.24\linewidth} p{0.40\linewidth}}
				\textbf{Term} & \textbf{Name/type} & \textbf{Physical effect (origin)} \\
				\midrule
				$\displaystyle k_{\perp}^{2}-k_{z}^{2}\omega_{2}^{2} +2\beta_{B}(\ell-\phi-k_{z}\omega_{1})$ &
				Constant shift &
				Defines the energy baseline, combining the geometric shift from the twist ($\omega_{2}$) and the Landau shift (dependent on `B`). \\
				\addlinespace
				$\displaystyle -\frac{(\ell-\phi-k_{z}\omega_{1})^{2}}{r^{2}}$ &
				Centrifugal/AB barrier &
				Governs the $r\!\to\!0$ behaviour via the effective index $\nu=|\ell-\phi-k_{z}\omega_{1}|$, which now includes the magnetic AB flux $\phi$. \\
				\addlinespace
				$\displaystyle +\frac{2k_{z}\omega_{2}\,(\ell-\phi-k_{z}\omega_{1})}{r}$ &
				Coulomb-like term &
				Mixes radial and azimuthal motion, induced by the local twist $\omega_{2}$ and axial momentum $k_z$. \\
				\addlinespace
				$\displaystyle - 2k_{z}\omega_{2}\,\beta_{B}\,r$ &
				Linear tilt (interaction) &
				Interaction term between the twist ($\omega_{2}$) and the magnetic field ($B$). Tilts the confinement potential, present only when $\omega_2, B \neq 0$. \\
				\addlinespace
				$\displaystyle - \beta_{B}^{2}\,r^{2}$ &
				Landau confinement &
				Parabolic confinement potential due to the uniform magnetic field $B$, leading to Landau levels. \\
			\end{tabular}
		\end{ruledtabular}
	\end{table}
	
	Table~\ref{tab:radial_terms_EM} extends this decomposition to the electromagnetic case (Eq.~(\ref{eq:radial_EM_expanded})). The Landau contribution $-\beta_B^2 r^2$ produces
	the familiar parabolic confinement, while the linear term
	$-2k_z\omega_2\beta_B\,r$—present only when both $B\neq0$ and $\omega_2\neq0$—tilts the effective well and shifts weight radially, explaining the downshift of $E_n$ with
	increasing $\omega_2$ at fixed $|B|$. The AB shift enters as
	$\ell\to\ell-\phi-k_z\omega_1$, preserving periodicity in $\phi$, and the Coulomb-like $1/r$ mixing remains the key local geometric signature. Together, these terms account for the Landau-fan trends, avoided crossings, and current profiles
	reported later.
	
	\noindent\emph{Remarks for interpretation and numerics.} The near-axis behaviour is governed by the effective index $\nu=|\ell-\phi-k_{z}\omega_{1}|$, which fixes the Frobenius exponent and the local scaling; in numerics we impose a regular-core boundary condition via Dirichlet at a small $r_{\min}$. With $B\neq0$ and $\omega_{2}\neq0$, the linear term $-2k_{z}\omega_{2}\beta_{B}\,r$ introduces an asymmetric tilt of the Landau well, shifting probability weight and lowering energies for electrons with $B>0$ (since $\beta_B<0$). These features underlie the level trends and current profiles discussed elsewhere; they also provide clear experimental handles: AB oscillations in $\phi$, reindexing with period $\Delta\omega_{1}=1/k_{z}$, and a roughly linear decrease of $E_n$ with $\omega_{2}$ at small twist due to geometric mixing.
	
	\textbf{Reader's guide.}
	The geometry used in this work can be read with three simple ideas in mind.
	(i) The function $f(r)=\omega_{1}+\omega_{2}r$ mixes rotations and translations, acting as a geometric gauge potential in the substitution (\ref{sub}) (see Eqs.~(\ref{eq:radialpiece})-(\ref{eq:angz})).
	(ii) The constant piece $\omega_{1}$ is ``global'': it shifts holonomies like an AB flux and reindexes $\ell\!\to\!\ell-k_{z}\omega_{1}$ without generating local curvature.
	(iii) The linear piece $\omega_{2}r$ is ``local'': it produces $de^{z}=\omega_{2}\,dr\wedge d\varphi$ and hence curvature, leading to $1/r$ mixing and a position-dependent coupling in the quantum problem.
	These points explain why $\omega_{1}$ produces AB-like periodicities, whereas $\omega_{2}$ tilts and mixes the radial dynamics even at fixed $k_{z}$.
	
	\vspace{0.6em}
	
	\begin{figure}[!t]
		\centering
		\begin{tikzpicture}[scale=1.0]
			\begin{scope}[opacity=0.12]
				\draw[fill=black] (-2.2,0) ellipse (2.1 and 0.45);
				\draw (-2.2,0) ellipse (2.1 and 0.45);
				\draw (-2.2,4.0) ellipse (2.1 and 0.45);
				\draw (-4.3,0) -- (-4.3,4.0);
				\draw ( -0.1,0) -- ( -0.1,4.0);
			\end{scope}
			
			\def\rone{1.1}
			\def\rtwo{1.7}
			\draw[thick] (-2.2,0) circle (\rone);
			\draw[thick] (-2.2,0) circle (\rtwo);
			
			\draw[very thick, dash dot] 
			plot [domain=0:290, samples=120] 
			({-2.2 + \rtwo*cos(\x)}, {0.015*\x + \rtwo*0.26*sin(\x)})
			node[above right] {};
			
			\pgfmathsetmacro\ang{35}
			\coordinate (P) at ({-2.2 + \rtwo*cos(\ang)}, {\rtwo*sin(\ang)});
			\draw[->,thick] (P) -- ++({cos(\ang)},{sin(\ang)}) node[above right] {$e^{r}=dr$};
			\draw[->,thick] (P) -- ++({-sin(\ang)},{cos(\ang)}) node[above left] {$e^{\varphi}=r\,d\varphi$};
			\draw[->,thick] (P) -- ++({-0.9*sin(\ang)},{0.9*cos(\ang)+0.9})
			node[right] {$e^{z}=dz+f(r)\,d\varphi$};
			
			\draw[dashed] (-2.2,0) -- ++({\rtwo*cos(\ang)},{\rtwo*sin(\ang)});
			\draw[dashed] (-2.2,0) -- ++({\rone*cos(\ang)},{\rone*sin(\ang)});
			\node[fill=white, inner sep=1pt] at ({-2.2 + 0.55*cos(\ang)},{0.55*sin(\ang)}) {$r_{1}$};
			\node[fill=white, inner sep=1pt] at ({-2.2 + 0.85*\rtwo*cos(\ang)},{0.85*\rtwo*sin(\ang)}) {$r_{2}$};
			
			\node[fill=white, inner sep=2pt, align=left] at (1.6,2.65)
			{$f(r)=\omega_{1}+\omega_{2}r$\\[2pt]
				$de^{z}=\omega_{2}\,dr\wedge d\varphi$};
			
			\node[fill=white, inner sep=2pt] at (1.45,0.35) {helical lift of $e^{z}$};
		\end{tikzpicture}
		\caption{Geometric coframe for the twisted screw background. The 1-forms are
			$e^{r}=dr$, $e^{\varphi}=r\,d\varphi$, and $e^{z}=dz+f(r)\,d\varphi$ with
			$f(r)=\omega_{1}+\omega_{2}r$. The constant part $\omega_{1}$ generates a
			holonomy (AB-like reindexing) without local density, whereas $\omega_{2}$ yields
			$de^{z}=\omega_{2}\,dr\wedge d\varphi$ and an $r$-dependent mixing of rotations
			and translations. The dashed-dot curve sketches the helical lift of $e^{z}$
			along a circular path at radius $r_{2}$; two radii $r_{1}<r_{2}$ illustrate the
			radius dependence of the geometric phase.}
		\label{fig:coframe-helix}
	\end{figure}
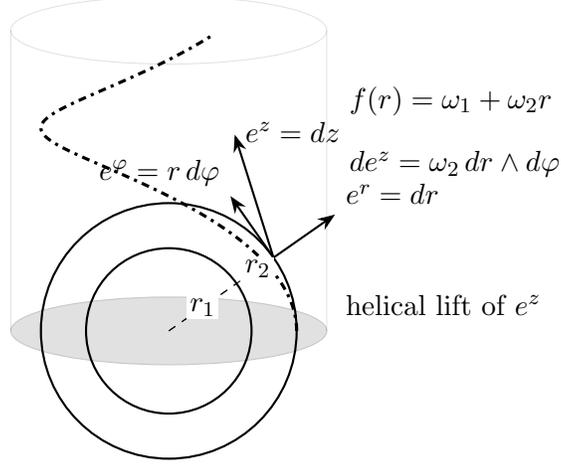
	Figure~\ref{fig:coframe-helix} visualizes the geometric coframe $(e^{r},e^{\varphi},e^{z})$ introduced in Sec.~\ref{sec:geometry}. The 1-forms $e^{r}\!=dr$, $e^{\varphi}\!=r\,d\varphi$, and $e^{z}\!=dz+f(r)\,d\varphi$ (with $f(r)=\omega_{1}+\omega_{2}r$, cf. Eq.~(\ref{fr})) make explicit the \emph{helical lift} of the $z$-direction along circular orbits. The dashed-dot curve sketches how $e^{z}$ acquires an azimuthal component set by $f(r)$, while the two radii $r_{1}<r_{2}$ emphasize that only the $\omega_{2}$-part generates a radius-dependent geometric “field’’ ($de^{z}=\omega_{2}\,dr\wedge d\varphi$). This picture underlies the effective substitution (\ref{sub}) in the Laplace-Beltrami operator (Eq.~(\ref{sub})) and the geometric phase per turn $\Delta\Phi=2\pi k_{z}(\omega_{1}+\omega_{2}r)$ (see Sec.~\ref{sec:phase}): $\omega_{1}$ acts as a radius-independent screw “flux’’ (AB-like), whereas $\omega_{2}$ imprints an $r$-dependent coupling that mixes azimuthal and axial motion.
	
	\section{Probability Current: Continuity, Azimuthal and Axial Components}
	\label{sec:currents}
	
	Here we collect the continuity law and the explicit expressions used to compute the probability currents. With
	$\rho=|\Psi|^{2}$ and $\sqrt{\gamma}=r$, probability is locally conserved,
	\begin{equation}
		\partial_{t}\rho \;+\; \frac{1}{\sqrt{\gamma}}\,
		\partial_{i}\!\big(\sqrt{\gamma}\,j^{i}\big)=0,
		\qquad
		j^{i}=\frac{\hbar}{m^{\ast}}\,
		\mathrm{Im}\!\big(\Psi^{\!*}\,\gamma^{ij}D_{j}\Psi\big),
		\quad i\in\{r,\varphi,z\}.
		\label{eq:cont-law}
	\end{equation}
	For stationary modes $\Psi(r,\varphi,z)=e^{ik_{z}z}e^{i\ell\varphi}R_{\ell}(r)$ and in the gauge
	$A_{r}=A_{z}=0$ with $\tfrac{q}{\hbar}A_{\varphi}(r)=\phi+\beta_{B}r^{2}$,
	the current components take the closed forms (using Eq. (\ref{fr}))
	\begin{equation}
		\label{eq:currents_components}
		\begin{aligned}
			j^{\varphi}(r)
			&= \frac{\hbar}{m^{\ast}}\,
			\frac{\ell-\phi-\beta_{B}r^{2}-k_{z}f(r)}{r^{2}}\;|\Psi|^{2},\\
			j^{z}(r)
			&= \frac{\hbar}{m^{\ast}}\!\left[
			\left(1+\frac{f(r)^{2}}{r^{2}}\right)k_{z}
			-\frac{f(r)}{r^{2}}\bigl(\ell-\phi-\beta_{B}r^{2}\bigr)
			\right]\!|\Psi|^{2}.
		\end{aligned}
	\end{equation}
	In the numerics we work with the Langer field $u=\sqrt{r}\,R_{\ell}$, so that
	$|\Psi|^{2}=|u|^{2}/r$; this choice preserves the discrete version of
	Eq.~(\ref{eq:cont-law}).
	
	The structure of Eq.~(\ref{eq:currents_components}) makes clear how each control parameter acts.
	The global twist $\omega_{1}$ enters as an AB-like shift $\ell\!\to\!\ell-k_{z}\omega_{1}$.
	The local twist $\omega_{2}$ produces an $r$-dependent mixing between azimuthal and axial
	transport through $f(r)$. A uniform magnetic field contributes via
	$\beta_{B}=qB/(2\hbar)$, adding the Landau term $\propto r^{2}$.
	The AB flux shifts $\ell\to\ell-\phi$, which implies periodicity
	$\Delta\phi=1$ in $j^{\varphi}$.
	
	Close to the axis, the trends simplify further: as $r\!\to\!0$, the azimuthal current is governed by
	$\ell-\phi-k_{z}\omega_{1}$, whereas the axial component is dominated by $k_{z}$. A negative $j^{z}$ very near the core can therefore appear whenever the screw coupling $f(r)$ overcompensates the $+k_{z}$ term; this backflow-like effect is physical.
	Its weight is small because the density behaves as $|\Psi|^{2}\sim r^{2\nu-1}$ with 
	$\nu=|\ell-\phi-k_{z}\omega_{1}|$.
	
	\begin{figure*}[tbhp]
		\centering
		\includegraphics[width=0.4\linewidth]{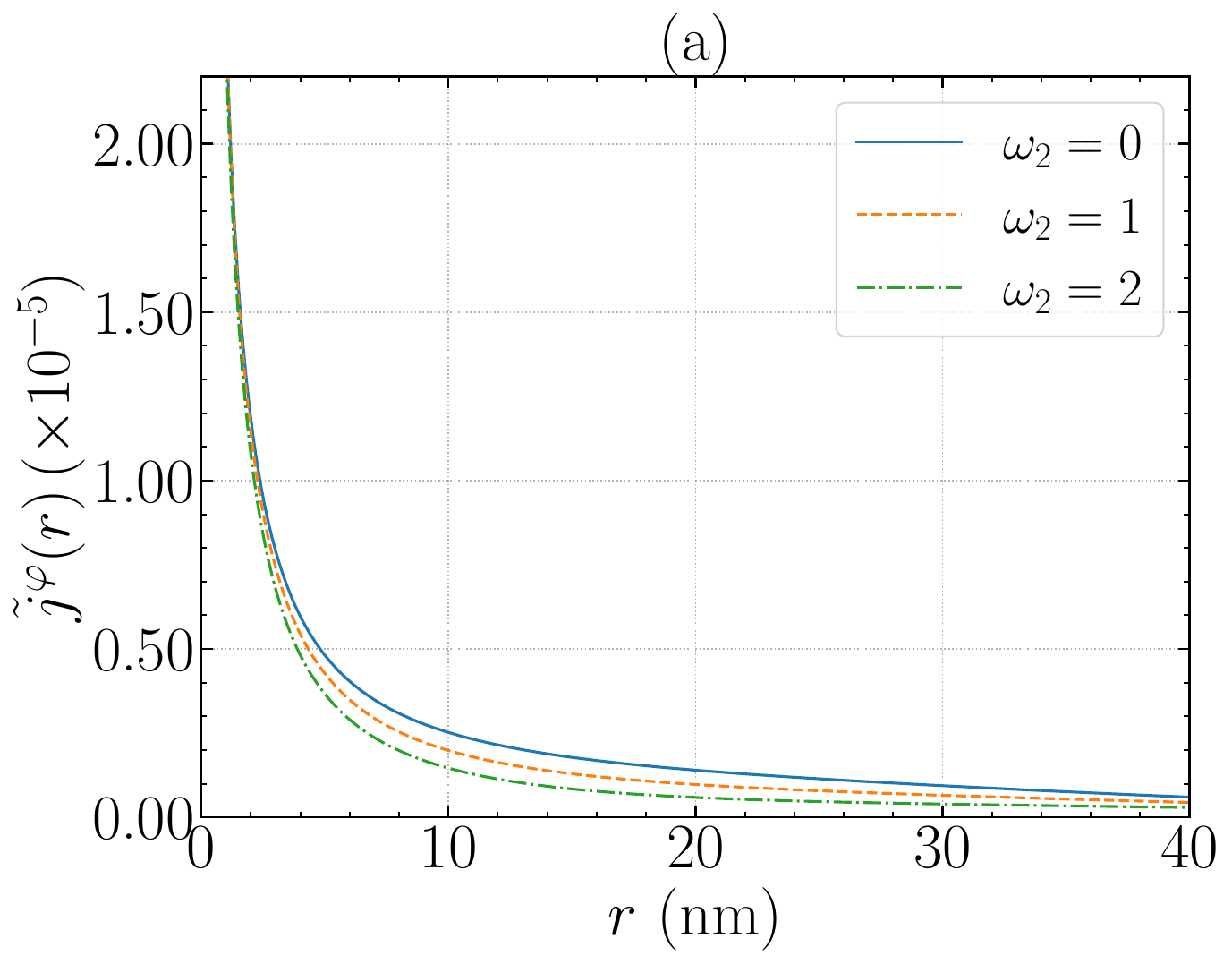}\qquad
		\includegraphics[width=0.4\linewidth]{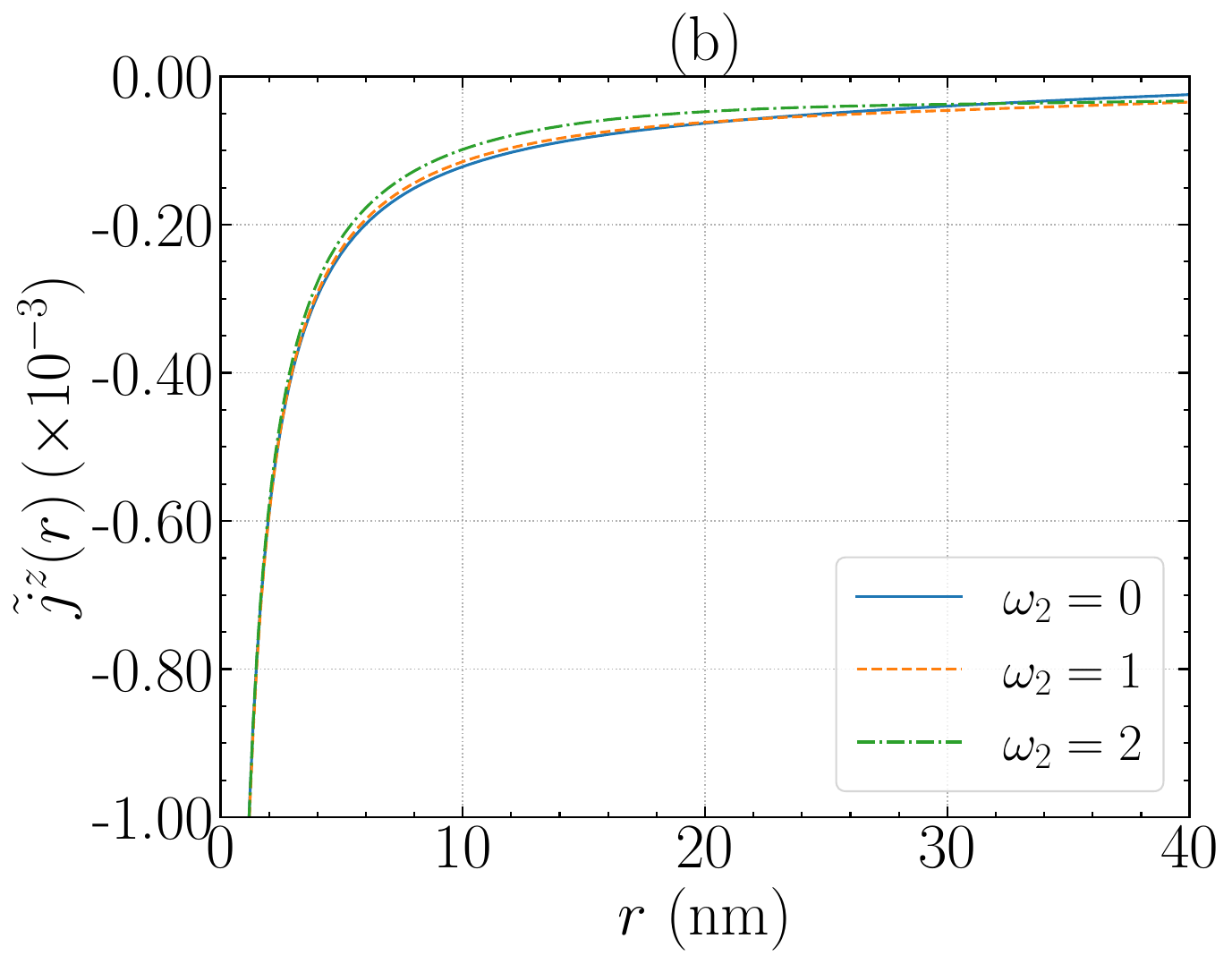}
		\caption{(a) Reduced azimuthal current $\tilde{j}^{\varphi}(r)$ and (b) reduced axial current $\tilde{j}^{z}(r)$ for the twisted-screw geometry.
			Curves correspond to $\omega_{2}\in\{0,1,2\}$ with the benchmark set $m^{\ast}/m_{e}=0.067$, $\ell=1$, $k_{z}=0.01~\mathrm{nm}^{-1}$, $\omega_{1}=50~\mathrm{nm}$, $\phi=0$, and $B=1~\mathrm{T}$ (electron).
			In (a) the azimuthal current decays rapidly with $r$ and its magnitude is slightly suppressed as $\omega_{2}$ increases, consistent with the $\omega_{2}$-induced tilt of the effective confinement which redistributes $|\Psi|^{2}$. In (b) the axial current is negative near the core and tends to $0$ from below; the position of its steep rise shifts mildly with $\omega_{2}$ through the twist-field coupling $\propto\omega_{2}\beta_{B}$. Both panels use the same plotting window, $r\in[0,40]~\mathrm{nm}$.}
		\label{fig:currents}
	\end{figure*}
	
	In practice, all profiles are obtained by solving the Langer Sturm-Liouville problem with Dirichlet boundaries on $r\in[r_{\min},r_{\max}]$, normalising $\int|u|^{2}\,dr=1$,
	and then evaluating the right-hand sides of Eq.~(\ref{eq:currents_components}).
	This guarantees consistency between the current profiles and the eigenstates and enforces
	Eq.~(\ref{eq:cont-law}) on the discrete grid.
	
	Figure~\ref{fig:currents} displays the radial profiles of the azimuthal and axial probability currents obtained from Eqs.~(\ref{eq:currents_components}) using the numerically computed eigenstates. Figure \ref{fig:currents}(a) shows that $j^{\varphi}(r)$ decays with radius and is mildly suppressed as the local twist $\omega_{2}$ increases, reflecting the $\omega_{2}$-induced redistribution of $|\Psi|^{2}$. Figure \ref{fig:currents}(b) highlights the near-axis \emph{backflow} in $j^{z}(r)$: for small $r$, the screw mixing via $f(r)$ can overcome the positive $k_{z}$ contribution and make $j^{z}<0$, while the profile recovers the sign of $k_{z}$ at larger radii. The position of the steep rise in $j^{z}$ shifts weakly with $\omega_{2}$ through the twist-field coupling (visible in the linear-in-$r$ term of the expanded radial equation), consistent with the trends discussed in Secs.~\ref{sec:num_methods_results} and \ref{sec:currents_reprise}.
	\begin{table}[t]
		\centering
		\caption{Control knobs and primary observables.}
		\label{tab:controls}
		\small
		\begin{tabular}{ll}
			\toprule
			Knob & Primary observable/signature \\
			\midrule
			$\omega_{1}$ & AB-like shift $\ell\!\to\!\ell-k_{z}\omega_{1}$; period $\Delta\omega_{1}=1/k_{z}$ in $E(\omega_{1})$ \\
			$\omega_{2}$ & $1/r$ mixing term; radius-dependent phase $2\pi k_{z}\omega_{2}(r_{1}-r_{2})$; level tilts/avoided crossings \\
			$B$ & Landau parabola ($-\beta_{B}^{2}r^{2}$) and linear tilt ($-2k_{z}\omega_{2}\beta_{B}r$) when $\omega_{2}\!\neq\!0$ \\
			$\Phi$ & AB oscillations: $\ell\!\to\!\ell-\phi$; currents $j_{\varphi}$ periodic in $\phi$ \\
			$k_{z}$ & Coupling strength to geometry; interferometric phase gain (\ref{eq:phase}) \\
			\bottomrule
		\end{tabular}
	\end{table}
	Table~\ref{tab:controls} compiles the control parameters and their primary experimental signatures together with how to disentangle them in practice. The global screw $\omega_{1}$ acts as an AB-like reindexing $\ell\!\to\!\ell-k_{z}\omega_{1}$, yielding periodicity $\Delta\omega_{1}=1/k_{z}$ in spectra and ring currents and cusp minima when $k_{z}\omega_{1}\!\approx\!\ell$ (ground-state relabeling). The local twist $\omega_{2}$ controls local mixing: it generates the $1/r$ term that couples radial and azimuthal motion and, when $B\!\neq\!0$, a linear-in-$r$ tilt $-2k_{z}\omega_{2}\beta_{B}r$ that shifts weight radially; observables include a roughly linear decrease of $E_{n}$ with $\omega_{2}$ at fixed $|B|$, a radius-dependent geometric phase $2\pi k_{z}\omega_{2}(r_{1}-r_{2})$ in interferometry, and a mild renormalization of the amplitude of $j^{\varphi}(r)$. A uniform field $B$ produces the Landau parabola ($-\beta_{B}^{2}r^{2}$) and a baseline shift $2\beta_{B}(\ell-\phi-k_{z}\omega_{1})$, giving fan-like $E_{n}(B)$ curves that are nearly even in $B$ (small odd-in-$B$ corrections appear only via the $\omega_{2}\beta_{B}$ mixing). The AB flux $\phi=\Phi/\Phi_{0}$ leads to periodicity $\Delta\phi=1$ in $E_{n}$ and in persistent currents, with cusp-like minima and $\ell\!\to\!\ell\pm1$ relabeling. Finally, $k_{z}$ sets the coupling strength to geometry and the phase sensitivity per revolution, scaling both the AB-like reindexing (period $1/k_{z}$) and the $\omega_{2}$ slopes (e.g., $\partial E_{n}/\partial\omega_{2}\propto k_{z}\beta_{B}$ at small $\omega_{2}$). Operationally, sweeping $\phi$ isolates pure AB oscillations, sweeping $\omega_{1}$ at fixed $k_{z}$ tests the reindexing period $1/k_{z}$, measuring $E_{n}$ versus $\omega_{2}$ at fixed $|B|$ quantifies local mixing, and reversing $B$ checks the sign of the linear tilt ($\propto k_{z}\omega_{2}\beta_{B}$), allowing each knob’s signature to be identified unambiguously.
	
	\subsection*{Numerical realization (discretization and solver)}
	
	To solve the radial problem in practice, we work with the Langer field $u(r)=\sqrt{r}\,R(r)$, which removes the first-derivative term and casts the equation into a symmetric Sturm-Liouville form suitable for stable discretization and standard boundary conditions. We write
	\begin{equation}
		-\,u''(r) + U(r)\,u(r) \;=\; \varepsilon\,u(r), 
		\qquad 
		U(r)=\frac{\bigl[\ell-\phi-k_{z}\bigl(\omega_{1}+\omega_{2}r\bigr)-\beta_{B}r^{2}\bigr]^{2}}{r^{2}}-\frac{1}{4r^{2}},
		\label{eq:Sl_U_en}
	\end{equation}
	with $E=\frac{\hbar^{2}}{2m^{\ast}}\bigl(\varepsilon+k_{z}^{2}\bigr)$. We impose Dirichlet boundary conditions on a sufficiently large interval $r\in[r_{\min},r_{\max}]$ and discretize the operator with a uniform second-order finite-difference grid. The kinetic part uses the standard three-point stencil, while $U(r)$ is assembled pointwise, including geometric, AB and uniform-field contributions. Lowest eigenpairs are obtained with a sparse shift-invert iteration; convergence is monitored by refining the mesh and expanding $r_{\max}$ until relative variations of the lowest levels fall below $10^{-3}$. Normalization is imposed on $u$, $\int |u|^{2}\,dr=1$, so that the configuration-space density used elsewhere is $|\Psi|^{2}=|u|^{2}/r$. Unless stated otherwise, we enforce a regular-core boundary condition via Dirichlet at a small $r_{\min}$ (and at a large $r_{\max}$), which is numerically robust under mesh and cutoff refinement.
	Unless stated otherwise, a mesoscopic benchmark compatible with semiconductor heterostructures is adopted: $m^{\ast}/m_{e}=0.067$, $\ell=1$, $k_{z}=0.01~\mathrm{nm}^{-1}$, $\omega_{1}=50~\mathrm{nm}$, $\phi=0$, and $B=1~\mathrm{T}$ for electrons (so $\beta_{B}=qB/2\hbar<0$).
	
	\section{Topological (Geometric) Phase}\label{sec:phase}
	
	The twisted-screw background introduces a natural notion of holonomy for matter waves, whereby a closed circuit around the axis produces a net displacement along $z$ and, consequently, a geometric phase. Unlike a conventional Berry phase—tied to adiabatic cycles in parameter space \cite{PRSLA.1984.392.45}, this phase is purely kinematic and arises from the spatial one-form that mixes rotations and translations. It therefore provides a clean diagnostic of how global (topological) and local (metric) aspects of the geometry enter quantum dynamics. In particular, the global screw parameter $\omega_{1}$ plays the role of a radius-independent “flux” that shifts angular quantum numbers in close analogy with the AB effect \cite{PR.1959.115.485}, whereas the distributed twist $\omega_{2}$ imprints a radius dependence that couples radial and azimuthal motion. The expressions derived below make this separation explicit and connect the accumulated phase to measurable interferometric signals (e.g., arm-radius contrast), establishing a direct route to read out $(\omega_{1},\omega_{2})$ from spectroscopy and transport.

	The one-form that encodes the helical structure,
	\begin{equation}
		\Theta^{z}\equiv dz+f(r)\,d\varphi, \qquad f(r)=\omega_{1}+\omega_{2}r,
	\end{equation}
	implies that, after one full revolution at fixed radius $r$, the axial coordinate accumulates a shift
	\begin{equation}
		\Delta z=\oint \! f(r)\,d\varphi = 2\pi f(r)=2\pi(\omega_{1}+\omega_{2}r).
	\end{equation}
	A matter wave with axial momentum $k_{z}$ therefore acquires the geometric phase
	\begin{equation}\label{eq:phase}
		\Delta\Phi = k_{z}\,\Delta z = 2\pi k_{z}\bigl(\omega_{1}+\omega_{2}r\bigr).
	\end{equation}
	This formula neatly separates a \emph{global} (topological) contribution, $\propto \omega_{1}$, from a \emph{local} (metric) one, $\propto \omega_{2}r$. The former is independent of $r$ and is the direct analogue of the AB phase: no local force is required, but the wave acquires a universal shift governed by an effective screw ``flux'' (the Burgers vector per turn). The latter makes the phase radius dependent; two circular paths at radii $r_{1}$ and $r_{2}$ pick up a relative phase $\Delta\Phi_{12}=2\pi k_{z}\omega_{2}(r_{1}-r_{2})$, which provides an interferometric handle to read out $\omega_{2}$ directly. From the Hamiltonian viewpoint, Eq.~(\ref{eq:radial}) embodies the same physics through the substitution $\ell\to \ell-k_{z}f(r)$: when $\omega_{2}=0$ this is a constant, AB-like shift of angular momentum, whereas for $\omega_{2}\neq0$ it becomes position dependent, coupling radial and angular motion and enabling radius-selective persistent currents. In practice, $\omega_{1}$ controls holonomy (reindexing and periodicities), while $\omega_{2}$ controls local mixing and level tilts, a dichotomy that will reappear throughout the numerical analysis.
	
	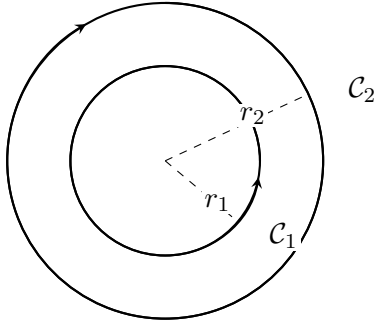
\begin{figure}[tbhp]
		\centering
		\begin{tikzpicture}[scale=1.05, >=stealth] 
			\draw[thick] (0,0) circle (2);
			\draw[thick] (0,0) circle (1.2);
			
			\draw[->,thick] (0,2) arc[start angle=90, end angle=-240, radius=2];
			\draw[->,thick] (1.2,0) arc[start angle=0, end angle=350, radius=1.2];
			
			\draw[dashed] (0,0) -- (25:2)
			node[midway, above right, fill=white, inner sep=1pt] {$r_{2}$};
			\draw[dashed] (0,0) -- (-40:1.2)
			node[midway, below right, fill=white, inner sep=1pt] {$r_{1}$};
			
			\node[fill=white, inner sep=1pt] at (2.5,0.9) {$\mathcal{C}_{2}$};
			\node[fill=white, inner sep=1pt] at (1.5,-0.95) {$\mathcal{C}_{1}$};
		\end{tikzpicture}
		\caption{Two circular paths $\mathcal{C}_{1}$ and $\mathcal{C}_{2}$ at radii $r_{1}$ and $r_{2}$. A wave with axial wavenumber $k_{z}$ accumulates the phase $\Delta\Phi=2\pi k_{z}(\omega_{1}+\omega_{2}r)$ per turn; the relative phase is $\Delta\Phi_{12}=2\pi k_{z}\omega_{2}(r_{1}-r_{2})$.}
		\label{fig:interferometer}
	\end{figure}
	
	As illustrated in Fig.~\ref{fig:interferometer}, two circular paths at radii $r_{1}$ and $r_{2}$ acquire a relative geometric phase
	$\Delta\Phi_{12}=2\pi k_{z}\omega_{2}(r_{1}-r_{2})$, which isolates the \emph{local} twist contribution: the global screw term $\propto\omega_{1}$ cancels out between the arms, leaving a contrast solely controlled by the radial separation $(r_{1}-r_{2})$, the axial wavenumber $k_{z}$, and the twist strength $\omega_{2}$. The sign of $\Delta\Phi_{12}$ flips with $k_{z}$ (or under $r_{1}\leftrightarrow r_{2}$), and the accumulated phase is independent of the AB flux, so that AB-induced reindexing affects both arms equally. In an interferometric readout, the fringe shift scales linearly with $\omega_{2}$ for small twists and with the arm-radius contrast, providing a direct metrological handle: maximizing $|r_{1}-r_{2}|$ and stabilizing the radial spread $\Delta r\ll |r_{1}-r_{2}|$ enhances visibility, while $k_{z}$ sets the phase sensitivity per revolution.
	
	\section{Numerical Methods and Results}\label{sec:num_methods_results}
	
	In this section, we solve the radial eigenproblem numerically by first applying the Langer transform $u(r)=\sqrt{r}\,R(r)$, which removes the first-derivative term and casts the equation into a symmetric Sturm--Liouville form on $(r_{\min},r_{\max})$, well suited to stable discretization and standard boundary conditions. The resulting operator is discretized on a uniform finite-difference grid $r\in[r_{\min},r_{\max}]$ with Dirichlet boundary conditions at both ends; the second derivative is represented by the standard three-point stencil and the effective potential is assembled pointwise, including geometric, AB, and uniform-field contributions. The lowest eigenpairs are obtained with a sparse shift-invert iteration; convergence is assessed by refining the mesh and enlarging $r_{\max}$ until relative changes in the lowest levels fall below $10^{-3}$. All parameter sweeps in $(\omega_{1},\omega_{2},B,\phi,k_{z})$ reuse the
	same grid and solver settings for consistency. The configuration-space density used in current evaluations is $|\Psi(r)|^{2}=|u(r)|^{2}/r$.
	
	To solve the radial equation (\ref{eq:Sl_U_en}), we impose Dirichlet boundaries on a sufficiently large interval $r\in[r_{\min},r_{\max}]$ and discretize with a second-order uniform finite-difference grid. Normalization is imposed directly on $u$, $\int_{r_{\min}}^{r_{\max}}\!|u|^{2}\,dr=1$, so that the probability density is $|\Psi|^{2}=|u|^{2}/r$. Unless stated otherwise, we adopt a mesoscopic benchmark compatible with semiconductor heterostructures:
	$m^{\ast}/m_{e}=0.067$, $\ell=1$, $k_{z}=0.01~\mathrm{nm}^{-1}$, $\omega_{1}=50~\mathrm{nm}$, $\phi=0$, and $B=1~\mathrm{T}$ with electron charge (hence
	$\beta_{B}=qB/2\hbar<0$). The typical grid uses $N=2000$ points on $r\in[10^{-3},\,500]~\mathrm{nm}$. Reported energies are
	\begin{equation}
		E \;=\; \frac{\hbar^{2}}{2m^{\ast}}\bigl(\varepsilon+k_{z}^{2}\bigr)
		\quad \text{(meV)}.
	\end{equation}
	
\begin{figure}[tbhp]
		\centering
		\includegraphics[width=0.4\linewidth]{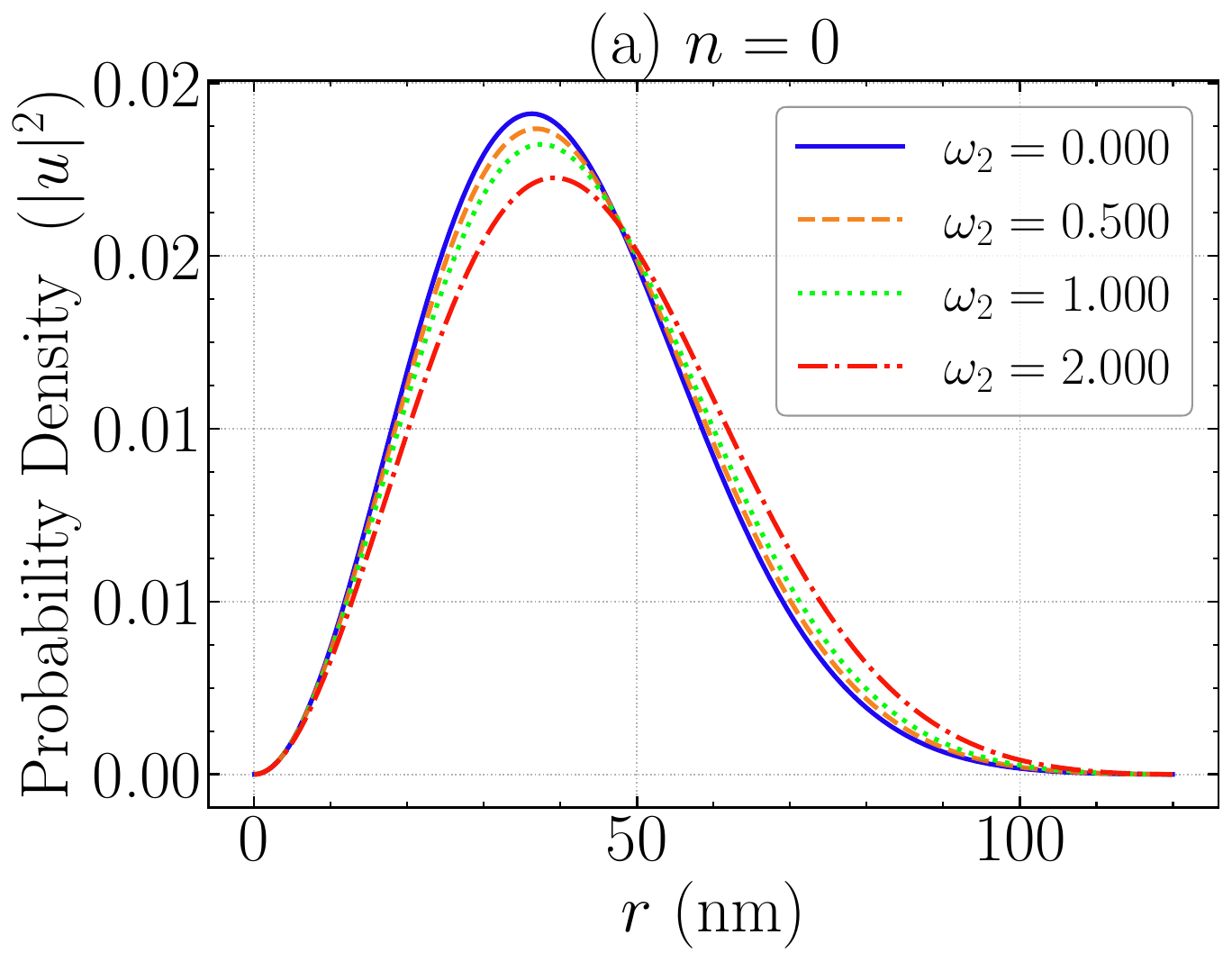}\qquad
		\includegraphics[width=0.4\linewidth]{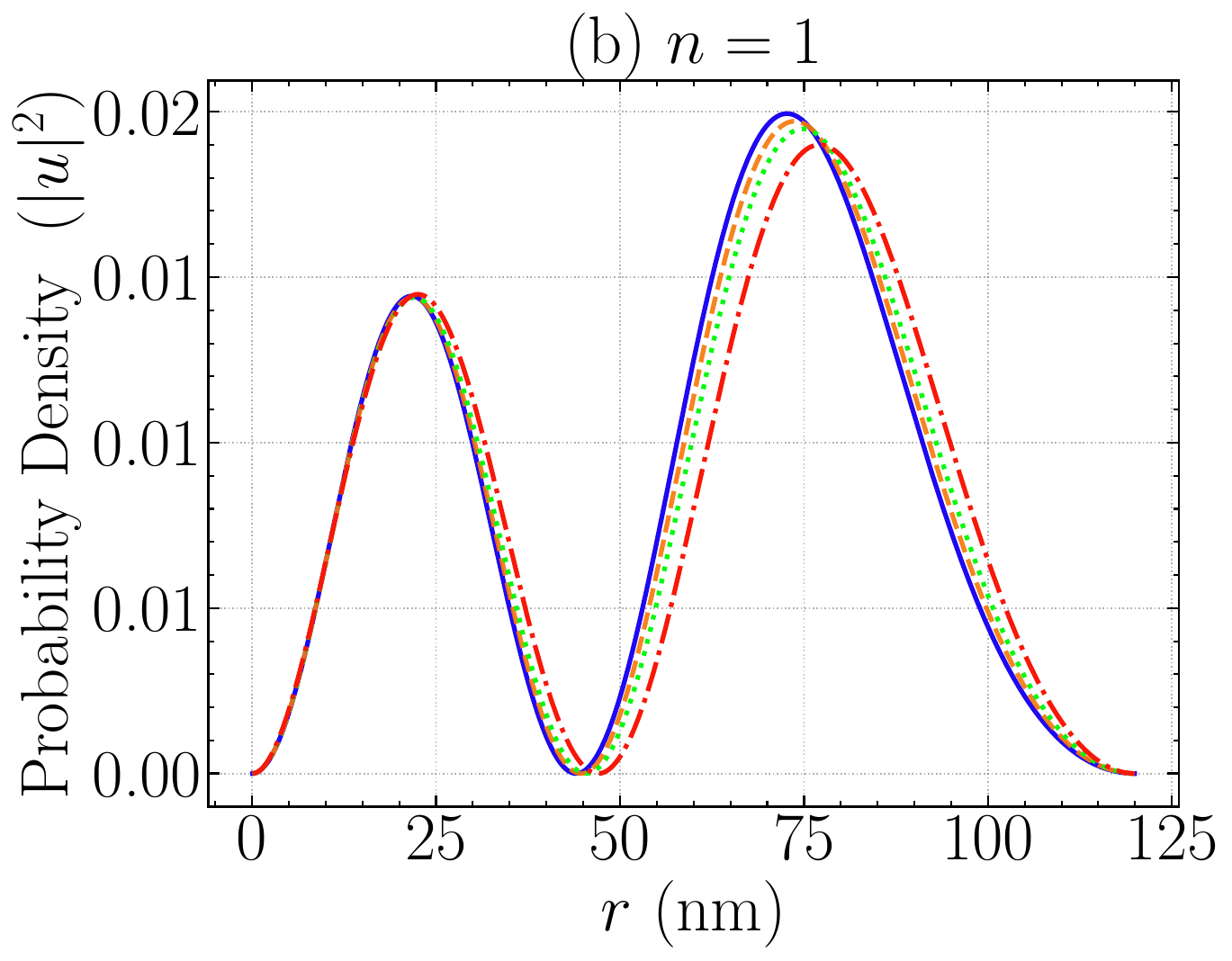}
		\caption{(a) and (b) Normalised radial probability densities $|u(r)|^{2}$ (with $\int |u|^{2}\,dr=1$ after the Langer transform $u=\sqrt{r}\,R$) for the ground state (a) and first excited state (b) in the twisted screw spacetime, for selected values of the local twist $\omega_{2}$ (see curve legends). Unless stated otherwise, parameters are $m^{\ast}/m_{e}=0.067$, $\ell=1$, $k_{z}=0.01~\mathrm{nm}^{-1}$, $\omega_{1}=50~\mathrm{nm}$, $\phi=0$ and $B=1~\mathrm{T}$ (electron, hence $\beta_{B}=qB/2\hbar<0$). Increasing $\omega_{2}$ tilts the effective well via the linear-in-$r$ term $-\,2\,k_{z}\,\omega_{2}\,\beta_{B}\,r$ in Eq.~(\ref{eq:radial_EM_expanded}), shifting probability weight and reducing $E_{n}$. Near the axis the behaviour is controlled by the centrifugal/AB barrier with index $\nu=|\ell-\phi-k_{z}\omega_{1}|$.}
		\label{fig:prob_u_panels}
	\end{figure}

	Figure~\ref{fig:prob_u_panels} shows the normalized radial densities $|u(r)|^{2}$ for the ground state (panel (a)) and first excited state (panel (b)) as the local twist
	$\omega_{2}$ is varied. Increasing $\omega_{2}$ \emph{tilts} the effective confinement and shifts probability weight, lowering the energy. This behavior follows directly from the linear-in-$r$ contribution,
	\begin{equation}
		-\,2\,k_{z}\,\omega_{2}\,\beta_{B}\,r,
	\end{equation}
	which appears upon expanding (\ref{eq:radial_EM_expanded}) when $B\neq0$ and
	$\omega_{2}\neq0$. For $B>0$ and electrons ($\beta_{B}<0$), this acts as a negative ramp
	in the effective potential $U(r)$, pushing levels downward as $\omega_{2}$ grows. The
	near-axis behavior remains governed by the centrifugal/AB barrier with
	effective index $\nu=|\ell-\phi-k_{z}\omega_{1}|$, so the nodal structure
	is unchanged: the first excited state has one internal zero while the ground state has
	none.
	
\begin{figure}[tbhp]
		\centering
		\includegraphics[width=0.6\linewidth]{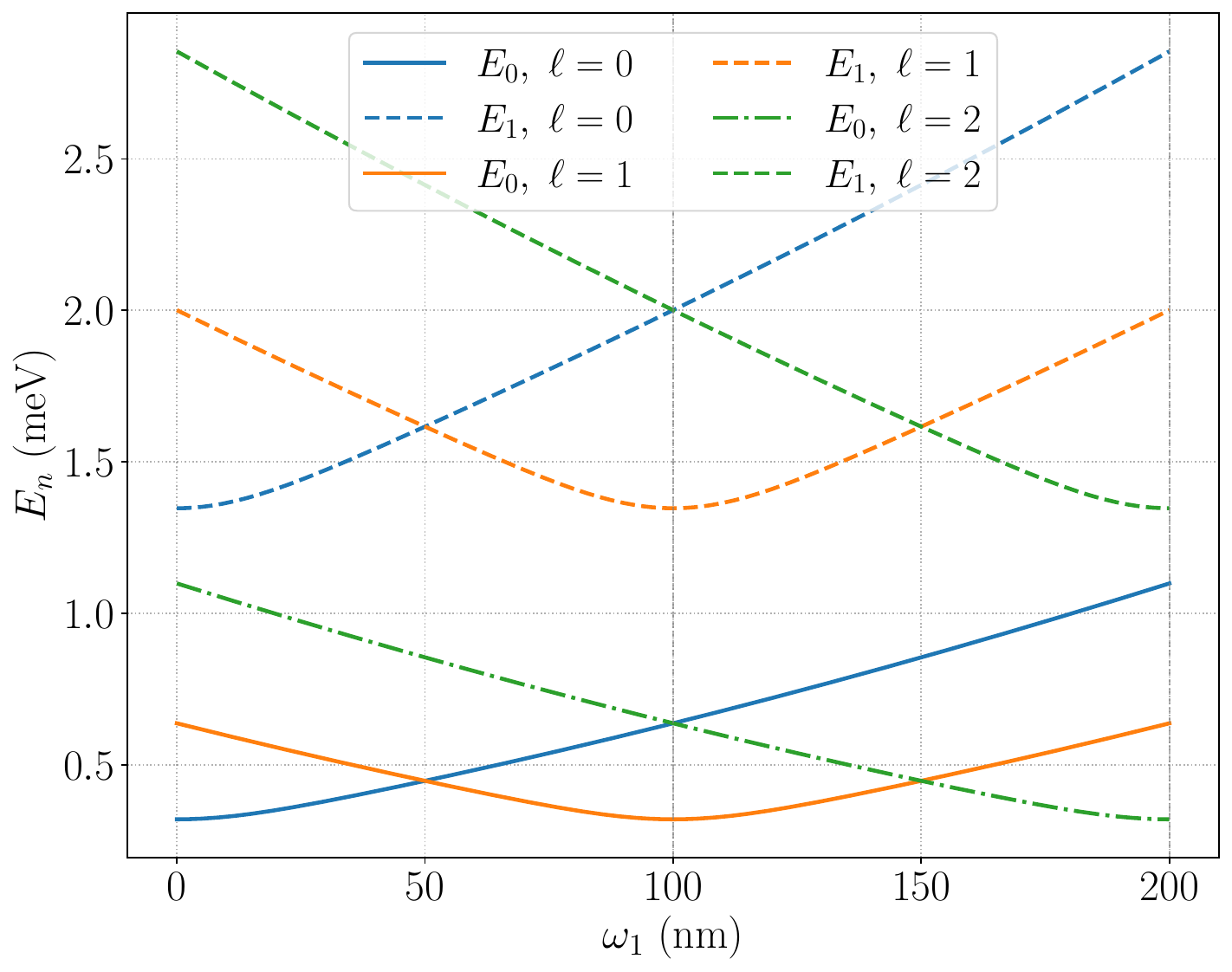}
		\caption{Energy levels $E_n$ as a function of the global screw parameter
			$\omega_{1}$ for three azimuthal indices $\ell\in\{0,1,2\}$ (GaAs parameters:
			$k_{z}=0.01~\mathrm{nm^{-1}}$, $\omega_{2}=0$, $B=0$, $\phi=0$).
			Vertical dashed lines mark multiples of the AB-like period
			$\Delta\omega_{1}=1/k_{z}=100~\mathrm{nm}$. 
			Each branch $E_{n}(\omega_{1};\ell)$ attains its minimum at
			$\omega_{1}\simeq \ell/k_{z}$ and maps onto the adjacent branch under the
			reindexing symmetry $E_{n}(\omega_{1}+\Delta\omega_{1};\ell)=E_{n}(\omega_{1};\ell-1)$,
			the hallmark of the global screw (AB-like) shift.}
		\label{fig:levels-omega1-multi-ell}
	\end{figure}

	The dependence on the global screw parameter $\omega_{1}$ is captured in Fig.~\ref{fig:levels-omega1-multi-ell}. In the absence of local twist and magnetic fields ($\omega_{2}=B=\phi=0$), the spectrum depends on $\omega_{1}$ only through the AB-like substitution $\ell\to\ell-k_{z}\omega_{1}$. Consequently, the centrifugal/AB index $\nu=|\ell-k_{z}\omega_{1}|$ controls the near-axis barrier and sets the energy scale. Each branch $E_{n}(\omega_{1};\ell)$ is minimized when $\nu$ is smallest, i.e. at $k_{z}\omega_{1}\approx \ell$, producing the parabolic envelopes with minima at $\omega_{1}\simeq \ell/k_{z}$. Because $\ell$ is integer, the spectrum has the reindexing (Floquet-like) symmetry $E_{n}(\omega_{1}+\Delta\omega_{1};\ell)=E_{n}(\omega_{1};\ell-1)$ with $\Delta\omega_{1}=1/k_{z}$, which manifests as the period marked by the vertical dashed lines. With $\omega_{2}=0$ and $B=0$ there is no coupling between different $\ell$, so crossings between branches belonging to different azimuthal indices are exact. Introducing a local twist ($\omega_{2}\neq 0$) or a magnetic field ($B\neq 0$) breaks this simple structure by mixing radial and angular motion: the minima shift, the curves tilt, and some of the crossings turn into avoided crossings (cf. Secs.~\ref{sec:schro} and \ref{sec:num_methods_results}).
	
\begin{figure}[tbhp]
		\centering
		\includegraphics[width=0.6\linewidth]{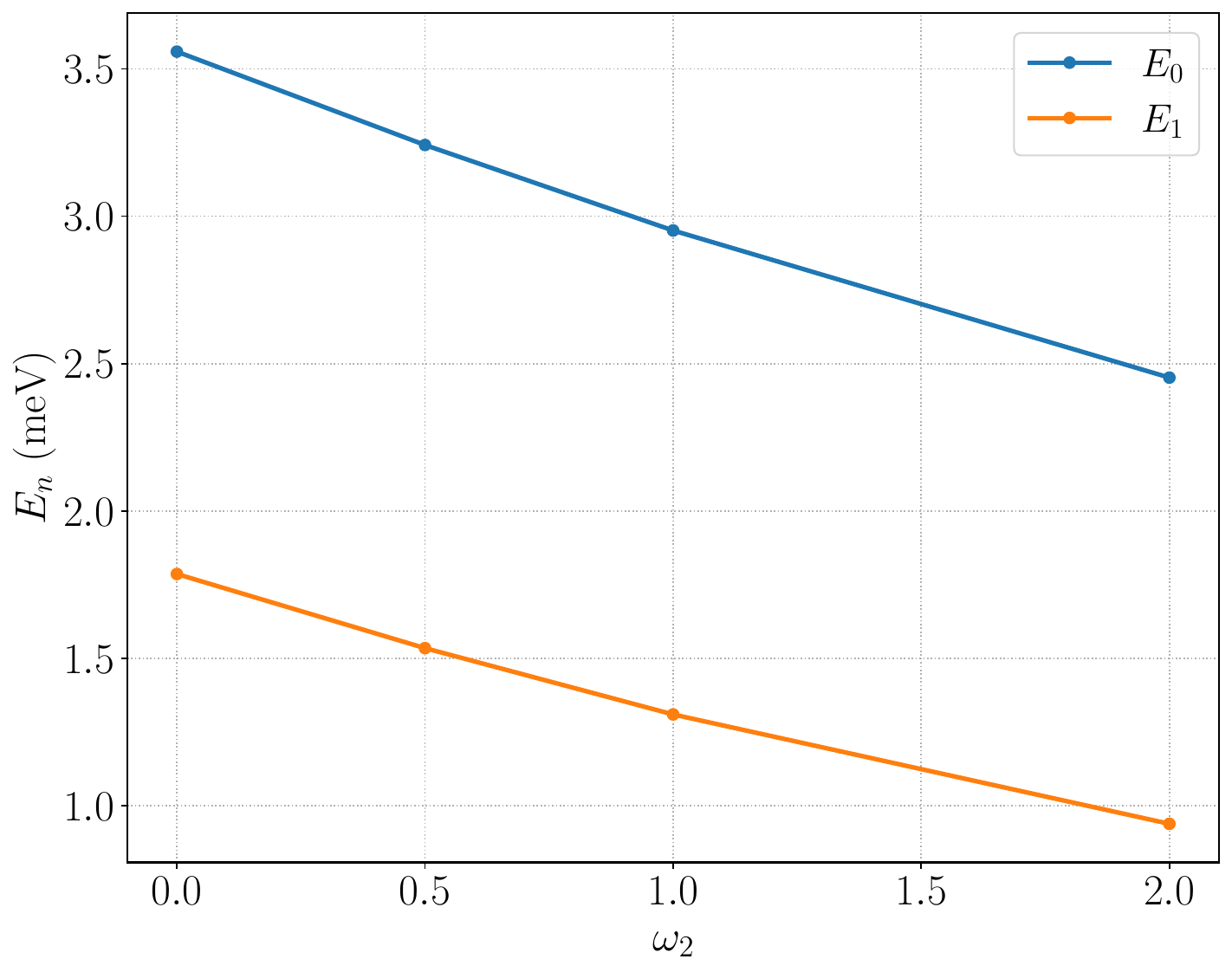}
		\caption{Lowest two eigenenergies $E_{0}$ and $E_{1}$ (meV) as functions of the local twist $\omega_{2}$ for the twisted screw metric. Unless stated otherwise, benchmark parameters are $m^{\ast}/m_{e}=0.067$, $\ell=1$, $k_{z}=0.01~\mathrm{nm}^{-1}$, $\omega_{1}=50~\mathrm{nm}$, $\phi=0$, and $B=1~\mathrm{T}$ (electron, hence $\beta_{B}=qB/2\hbar<0$). Increasing $\omega_{2}$ enhances the tilt of the Landau-like confinement via the linear term $-\,2\,k_{z}\,\omega_{2}\,\beta_{B}\,r$ in the expanded radial equation, yielding a monotonic decrease of $E_{n}$. For small $\omega_{2}$ the dependence is approximately linear, with weaker curvature from quadratic $\omega_{2}^{2}$ contributions.}
		\label{fig:levels_vs_omega2}
	\end{figure}

\begin{figure}[tbhp]
		\centering
		\includegraphics[width=0.6\linewidth]{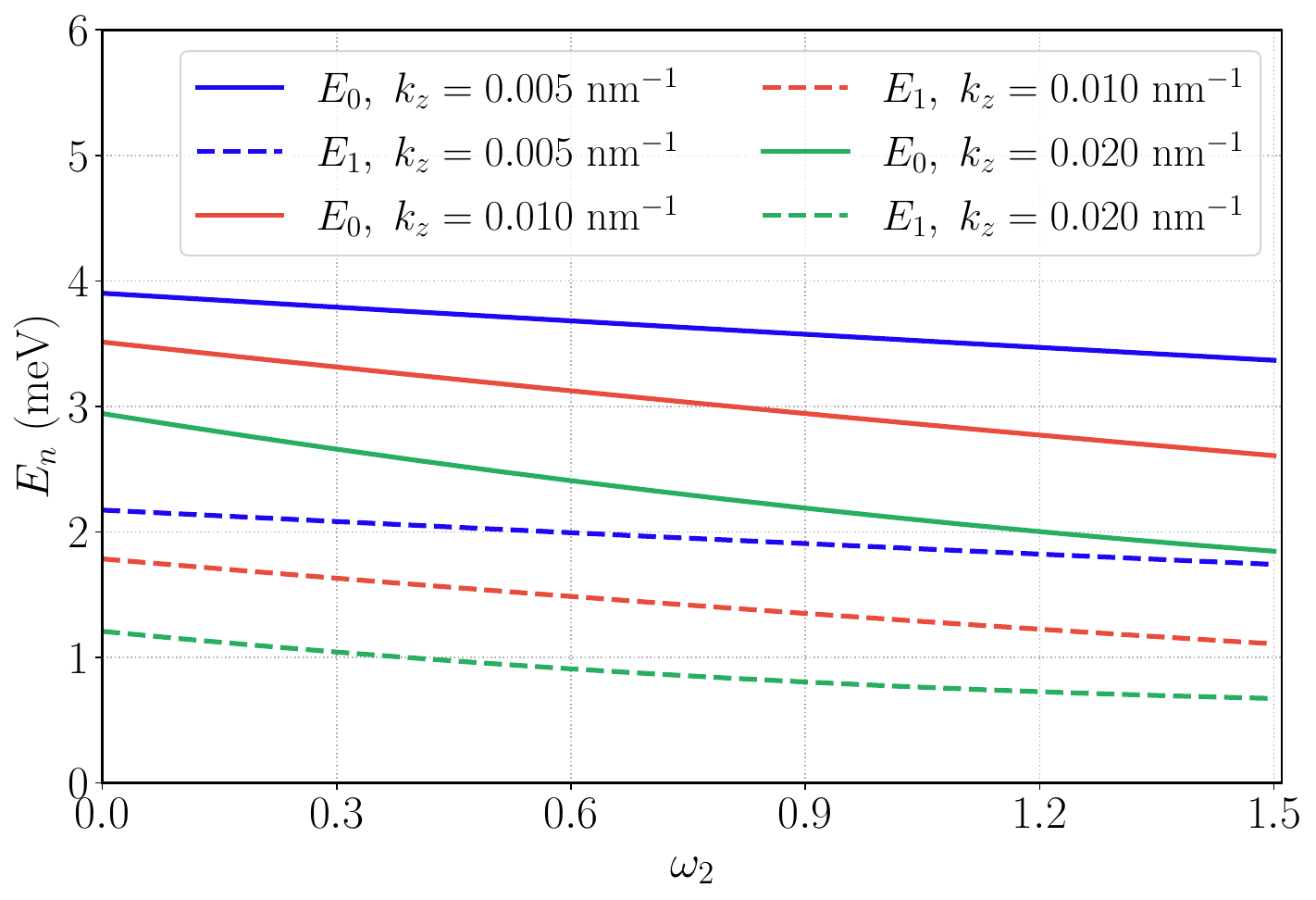}
		\caption{Lowest two eigenenergies $E_{0}$ (solid) and $E_{1}$ (dashed) as functions of the local twist $\omega_{2}$ for three axial wave numbers $k_{z}\in\{0.005,\,0.010,\,0.020\}\ \mathrm{nm^{-1}}$. Fixed parameters: $m^{\ast}/m_{e}=0.067$ (GaAs), $\ell=1$, $\omega_{1}=50\ \mathrm{nm}$, $\phi=0$, and $B=1\ \mathrm{T}$ for electrons (so $\beta_{B}=qB/2\hbar<0$). The range shown is $\omega_{2}\in[0,1.5]$. Increasing $\omega_{2}$ lowers the energies, with a slope that grows in magnitude with $k_{z}$, consistent with the linear-in-$r$ term $-\,2\,k_{z}\,\omega_{2}\,\beta_{B}\,r$ in the expanded radial equation. Slight deviations from perfect linearity reflect quadratic $\omega_{2}^{2}$ contributions and boundary effects.}
		\label{fig:levels_omega2_multi_kz}
	\end{figure}

	The role of the local twist $\omega_{2}$ is summarised in Figs.~\ref{fig:levels_vs_omega2} and \ref{fig:levels_omega2_multi_kz}. The roughly monotonic decrease of $E_{n}$ with $\omega_{2}$ is consistent with the interpretation above: a larger local twist strengthens the tilt of the Landau-like confinement through the coupling $\propto k_{z}\,\omega_{2}\,\beta_{B}$. For small $\omega_{2}$, the dependence is approximately linear, as expected from first-order perturbation theory about $\omega_{2}=0$. Quadratic terms in $\omega_{2}$, originating from the expansion of $\bigl[\ell-\phi-k_{z}(\omega_{1}+\omega_{2}r)-\beta_{B}r^{2}\bigr]^{2}$, produce a weaker curvature in $E_{n}(\omega_{2})$. The multi-$k_{z}$ plot further shows that the slope grows in magnitude with $k_{z}$, consistent with the linear-in-$r$ term $-\,2\,k_{z}\,\omega_{2}\,\beta_{B}\,r$; residual curvature reflects $\omega_{2}^{2}$ contributions and boundary effects. At $\omega_{2}=0$, the ordering and absolute scale reflect the AB-like shift $\ell\!\to\!\ell-k_{z}\omega_{1}$ and the Landau baseline; the near-axis behavior remains controlled by $\nu=|\ell-\phi-k_{z}\omega_{1}|$, so the nodal structure of $n=0$ and $n=1$ is unchanged across the sweep.
	
\begin{figure}[t]
		\centering
		\includegraphics[width=0.6\linewidth]{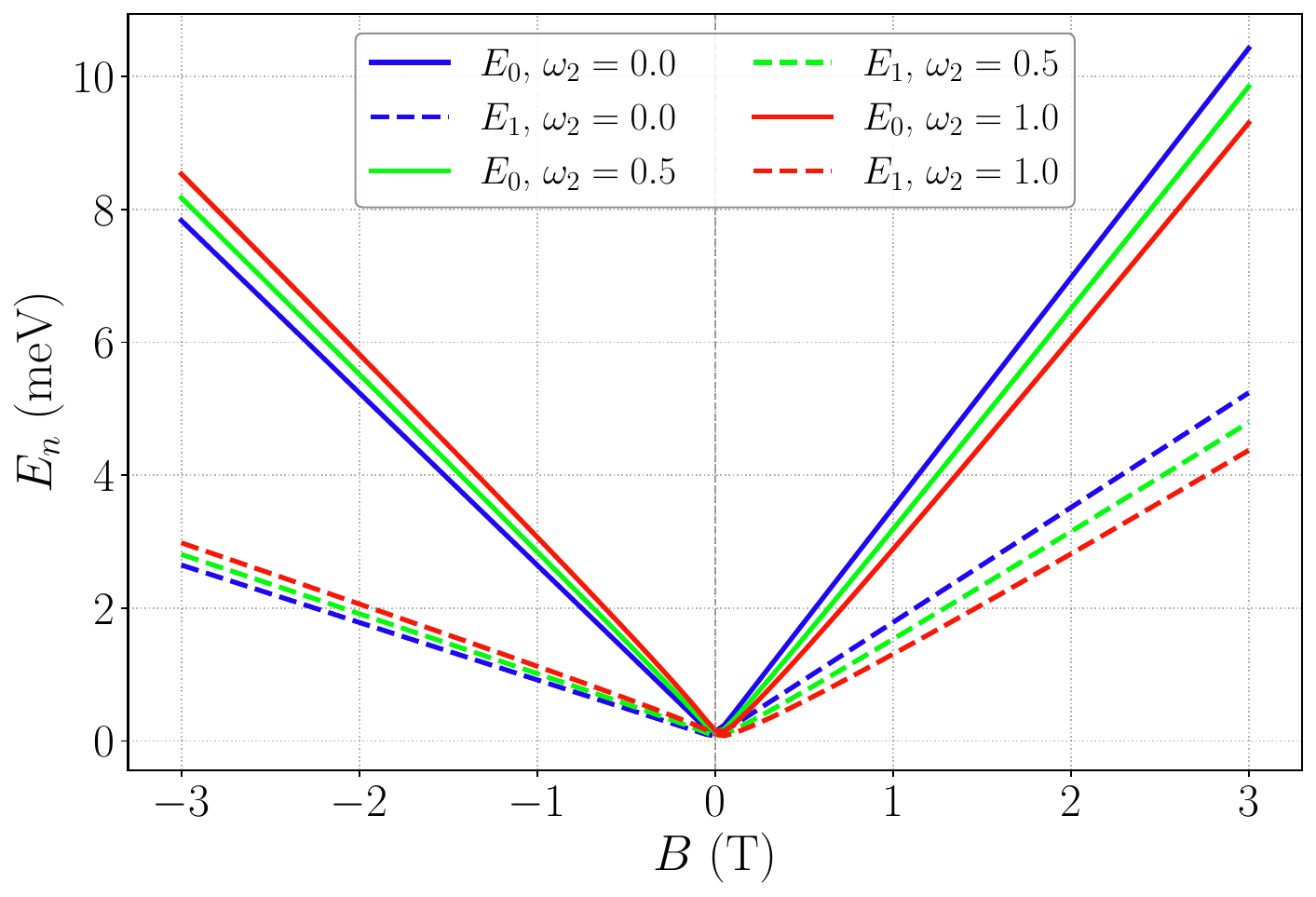}
		\caption{Lowest two eigenenergies $E_{0}$ (solid) and $E_{1}$ (dashed) as functions of the uniform magnetic field $B$ for three values of the local twist, $\omega_{2}\in\{0,\,0.5,\,1.0\}$ (colors as in the legend). Parameters held fixed: $m^{\ast}/m_{e}=0.067$, $\ell=1$, $k_{z}=0.01~\mathrm{nm}^{-1}$, $\omega_{1}=50~\mathrm{nm}$, and $\phi=0$. We take electron charge so that $\beta_{B}=qB/(2\hbar)<0$ for $B>0$. Numerical grid: $r\in[10^{-3},\,500]~\mathrm{nm}$ with $N=2000$ points, Dirichlet boundaries, Langer form $u=\sqrt{r}\,R$. Energies in meV. The branches rise approximately linearly with $|B|$ (Landau-like trend from the $-\beta_{B}^{2}r^{2}$ term in the effective potential), exhibiting a cusp at $B=0$. Increasing $\omega_{2}$ lowers both $E_{0}$ and $E_{1}$ at a given $|B|$, consistent with the twist-field mixing term $-\,2\,k_{z}\,\omega_{2}\,\beta_{B}\,r$ in the expanded radial equation, which effectively softens the radial confinement.}
		\label{fig:levels-vs-B-omega2}
	\end{figure}

	The effect of a uniform magnetic field is exhibited in Fig.~\ref{fig:levels-vs-B-omega2}, which shows the ``Landau fan'' $E_{n}(B)$ for the lowest two states while the global screw parameter and the AB flux are kept fixed. The dominant $B$--dependence comes from the harmonic contribution $-\beta_{B}^{2}r^{2}$ in the effective potential of the Langer equation~(\ref{eq:Sl_U_en}), producing a nearly even, $|B|$-like rise with a cusp at $B=0$. The local twist $\omega_{2}$ couples to the field through the linear term $-\,2\,k_{z}\,\omega_{2}\,\beta_{B}\,r$. For the parameters used here, this mixing reduces the net confinement and shifts the whole fan downward as $\omega_{2}$ increases. The effect is more visible for higher $n$, in line with the larger spatial extent of the excited states. Any odd-in-$B$ corrections induced by the linear term remain numerically small in the field range explored, so the curves are practically symmetric under $B\!\to\!-B$. These trends provide a direct way to read out the local twist from spectroscopy: at fixed $|B|$, $dE_{n}/d\omega_{2}<0$ and the magnitude of the shift grows with $n$ and with $k_{z}$, as anticipated by the structure of the effective potential.
	
\begin{figure}[t]
		\centering
		\includegraphics[width=0.60\linewidth]{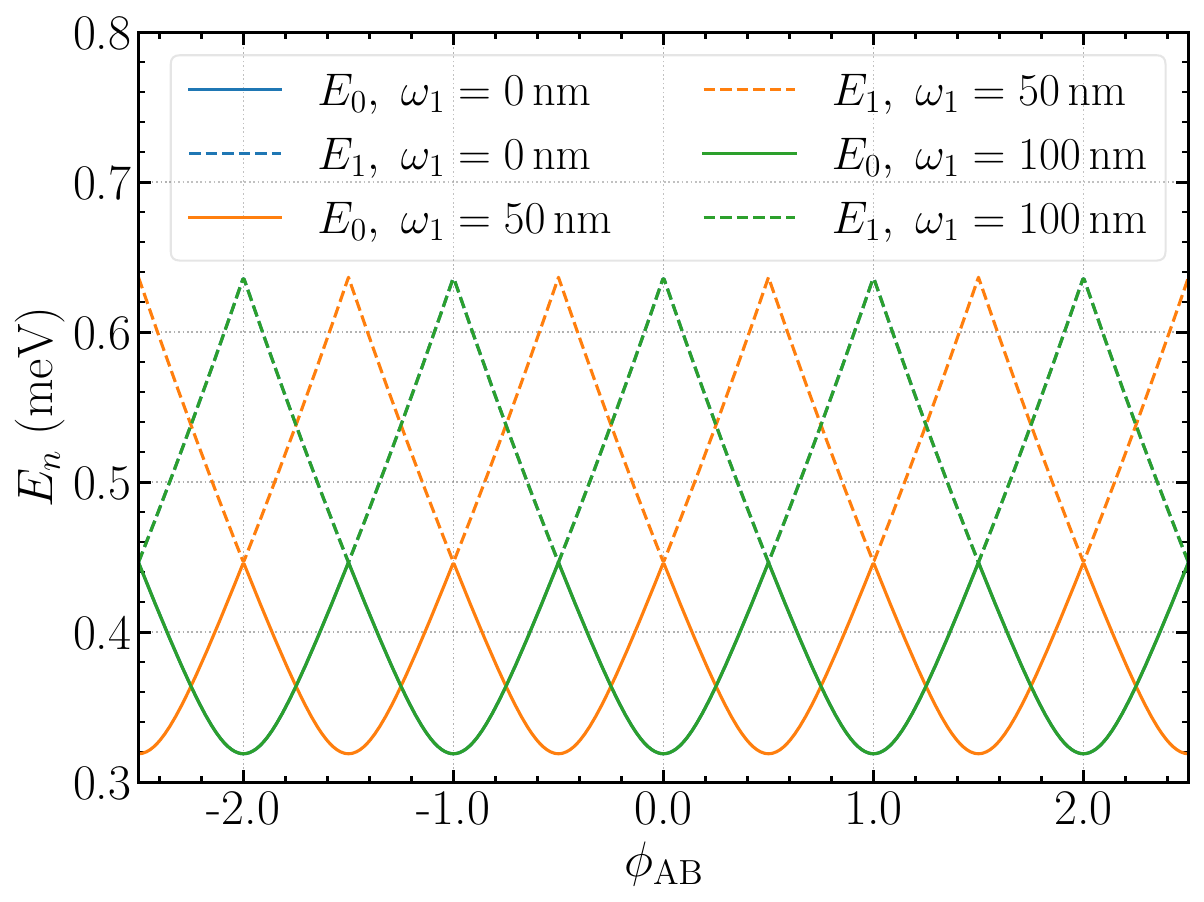}
		\caption{AB oscillations obtained from the full radial solver by taking, at each reduced flux $\phi=\Phi/\Phi_{0}$, the minimum over many azimuthal indices $\ell$ of the numerically computed levels $E_{n,\ell}(\phi)$.
			Thin, faint curves show representative branches $E_{0,\ell}$ (solid) and $E_{1,\ell}$ (dashed) for a few $\ell$'s; thick curves are the envelopes $E_{0}^{\rm env}(\phi)=\min_{\ell}E_{0,\ell}$ (solid) and $E_{1}^{\rm env}(\phi)=\min_{\ell}E_{1,\ell}$ (dashed).
			Colors correspond to three values of the global screw parameter $\omega_{1}\in\{0,50,100\}\,\mathrm{nm}$ (indicated in the legend).
			Parameters: $m^{\ast}/m_{e}=0.067$, $k_{z}=0.01~\mathrm{nm}^{-1}$, $\omega_{2}=0$, $B=0$.
			The envelopes display the expected AB periodicity $\Delta\phi=1$ with cusp-like minima at $\phi\simeq \text{integer}-k_{z}\omega_{1}$, evidencing ground-state transitions $ \ell\to\ell\pm1 $.
			The role of $\omega_{1}$ is a horizontal phase shift (AB-like), while the oscillation amplitude/curvature is essentially unchanged for fixed $k_{z}$ and material parameters.}
		\label{fig:AB-envelope}
	\end{figure}

	A complementary perspective comes from the AB oscillations extracted directly from the numerical spectrum in the field-free, homogeneous case ($\omega_{2}=0$, $B=0$). For each flux value $\phi=\Phi/\Phi_{0}$, we compute the lowest eigenenergies $E_{n,\ell}(\phi)$ for a large set of azimuthal indices $\ell$ and build the envelopes $E_{n}^{\rm env}(\phi)=\min_{\ell}E_{n,\ell}(\phi)$. The thin background curves in Fig.~\ref{fig:AB-envelope} are representative branches $E_{n,\ell}$ for a few $\ell$'s; the thick curves are the resulting envelopes for $n=0$ (solid) and $n=1$ (dashed). The envelopes exhibit the hallmark AB periodicity with period $\Delta\phi=1$, accompanied by cusp-like minima at $\phi\simeq \mathrm{integer}-k_{z}\omega_{1}$. These cusps signal ground-state relabeling events $\ell\to\ell\pm1$, i.e., the azimuthal quantum number that minimizes the energy changes by one unit at each flux quantum. Consistent with the geometric interpretation discussed earlier, the global screw parameter $\omega_{1}$ acts as an AB-like phase shifter: increasing $\omega_{1}$ simply translates the pattern along the horizontal axis by an amount $k_{z}\omega_{1}$, while the oscillation amplitude (curvature near the minima) remains essentially fixed for given $k_{z}$ and effective mass. This construction confirms that the oscillatory signal is not imposed by hand but emerges from the full radial problem by taking the minimum over many $\ell$ branches at each $\phi$.
	
	Finally, we collect limiting cases and consistency checks that are useful in practice and conceptually clarifying. With no magnetic field ($B=0$), the linear-in-$r$ term vanishes and the variation with $\omega_{2}$ is gentler, dominated by the $1/r$ mixing and by the $\propto \omega_{2}^{2}$ shift in $k_{\perp}^{2}$. For $k_{z}=0$, geometry-induced couplings disappear, recovering the standard forms (AB+Landau with field; Bessel/AB without field). With a finite AB flux, $\phi$ effectively shifts $\ell\to \ell-\phi$, and observables such as $E_{n}$ and azimuthal currents are periodic in $\phi$. The sign of $\beta_{B}$ matters: flipping it (holes instead of electrons, or reversing $B$) reverses the tilt and changes the slope $dE_{n}/d\omega_{2}$. For reproducibility, we verified numerical convergence by doubling the grid to $N=4000$ and enlarging $r_{\max}$ by $30\%$; the relative change in $E_{0}$ remains below $10^{-3}$. The cutoff $r_{\min}=10^{-3}$~nm ensures stability in the presence of $1/r$ and $1/r^{2}$ terms. The curves in Figs.~\ref{fig:prob_u_panels} and \ref{fig:levels_vs_omega2} were produced with the same parameter set and can be reproduced directly with the numerical routine described in the main text.
	
	\section{Azimuthal and Axial Probability Currents}\label{sec:currents_reprise}
	
	This section revisits the gauge-covariant probability current associated with the continuity equation introduced earlier, now specialized to the helical geometry and to the electromagnetic setting used throughout. The goal is to make explicit how each control parameter, global screw $\omega_{1}$, local twist $\omega_{2}$, uniform field $B$ and AB flux $\phi$, enters the current components, and to connect these analytic forms to the numerical profiles and integrated signals shown in the figures.
	
	Building on the covariant continuity law and the gauge-covariant derivatives introduced in Secs.~\ref{sec:schro} and \ref{sec:geometry}, and using the stationary ansatz $\Psi=e^{ik_{z}z}e^{i\ell\varphi}R_{\ell}(r)$ with $f(r)=\omega_{1}+\omega_{2}r$ and the vector potential of Eq.~(\ref{eq:gauge_phys_new}), the azimuthal and axial current components take the closed forms stated above. In all numerical evaluations, we work with the Langer field $u=\sqrt{r}\,R_{\ell}$, so that $|\Psi|^{2}=|u|^{2}/r$.
	
	The parameter dependence is transparent in those expressions: the global screw $\omega_{1}$ acts as an AB-like reindexing $\ell\!\to\!\ell-k_{z}\omega_{1}$; the local twist $\omega_{2}$ generates the $r$-dependent mixing that couples azimuthal and axial transport; a uniform field enters only through $\beta_{B}=qB/(2\hbar)$, adding the Landau term $\propto r^{2}$; and the AB flux shifts $\ell\!\to\!\ell-\phi_{\mathrm{AB}}$, leading to the standard periodicity $\Delta\phi_{\mathrm{AB}}=1$ in $j^{\varphi}$ at fixed values of the other controls.

	Near the axis, $r\to0$, the dominant factors are set by $\ell-\phi-k_{z}\omega_{1}$ and by $k_{z}$. In particular,
	\begin{equation}
		\operatorname{sgn}\,j^{z}(0^{+})
		=\operatorname{sgn}\!\left(\omega_{1}k_{z}-(\ell-\phi)\right),    
	\end{equation}
	so a \emph{negative} axial current in the core is not a numerical artifact, but a genuine backflow induced by the screw coupling: the term proportional to $f(r)$ can locally overcome the positive $k_{z}$ contribution. In the special case $\omega_{2}=0=B$, the zero of $j^{z}$ occurs at
	\begin{equation}
		r_{*}^{2}=\omega_{1}\,\frac{\ell-\phi}{k_{z}}-\omega_{1}^{2}    
	\end{equation}
	(whenever the right-hand side is positive), beyond which $j^{z}$ recovers the sign of $k_{z}$. Because the physical density scales as $|\Psi|^{2}\sim r^{2\nu-1}$ with $\nu=|\ell-\phi-k_{z}\omega_{1}|$, this near-axis backflow has small weight in cross-section integrals.
	
	For ring-like geometries, it is convenient to summarize the local information in line currents integrated over a narrow annulus around $r_{0}$, for example
	\begin{equation}
		I_{\varphi}(r_{0})=\int_{r_{0}-\delta}^{r_{0}+\delta} r\, j^{\varphi}(r)\,dr,
		\qquad
		I_{z}(r_{0})=\int_{r_{0}-\delta}^{r_{0}+\delta} r\, j^{z}(r)\,dr.    
	\end{equation}
	Combining Eqs. (\ref{eq:currents_components}) with the reindexing symmetry of the spectrum,
	$E_{n}(\omega_{1}+\Delta\omega_{1};\ell)=E_{n}(\omega_{1};\ell-1)$ where
	$\Delta\omega_{1}=1/k_{z}$ (see Sec.~\ref{sec:num_methods_results}),
	one obtains the same periodicity in $I_{\varphi}$ when sweeping either $\phi$ (period $1$) or $\omega_{1}$ (period $1/k_{z}$). When $\omega_{2}\neq0$ and $B\neq0$, the twist-field mixing term in the effective potential (Sec.~\ref{sec:schro}) smoothly shifts the radius at which $j^{z}$ attains its maximum by an amount controlled by $\omega_{2}\beta_{B}$; in practice, this shift tracks the radial redistribution of $|\Psi|^{2}$.
	\begin{figure}[tbhp]
		\centering
		\includegraphics[width=0.6\linewidth]{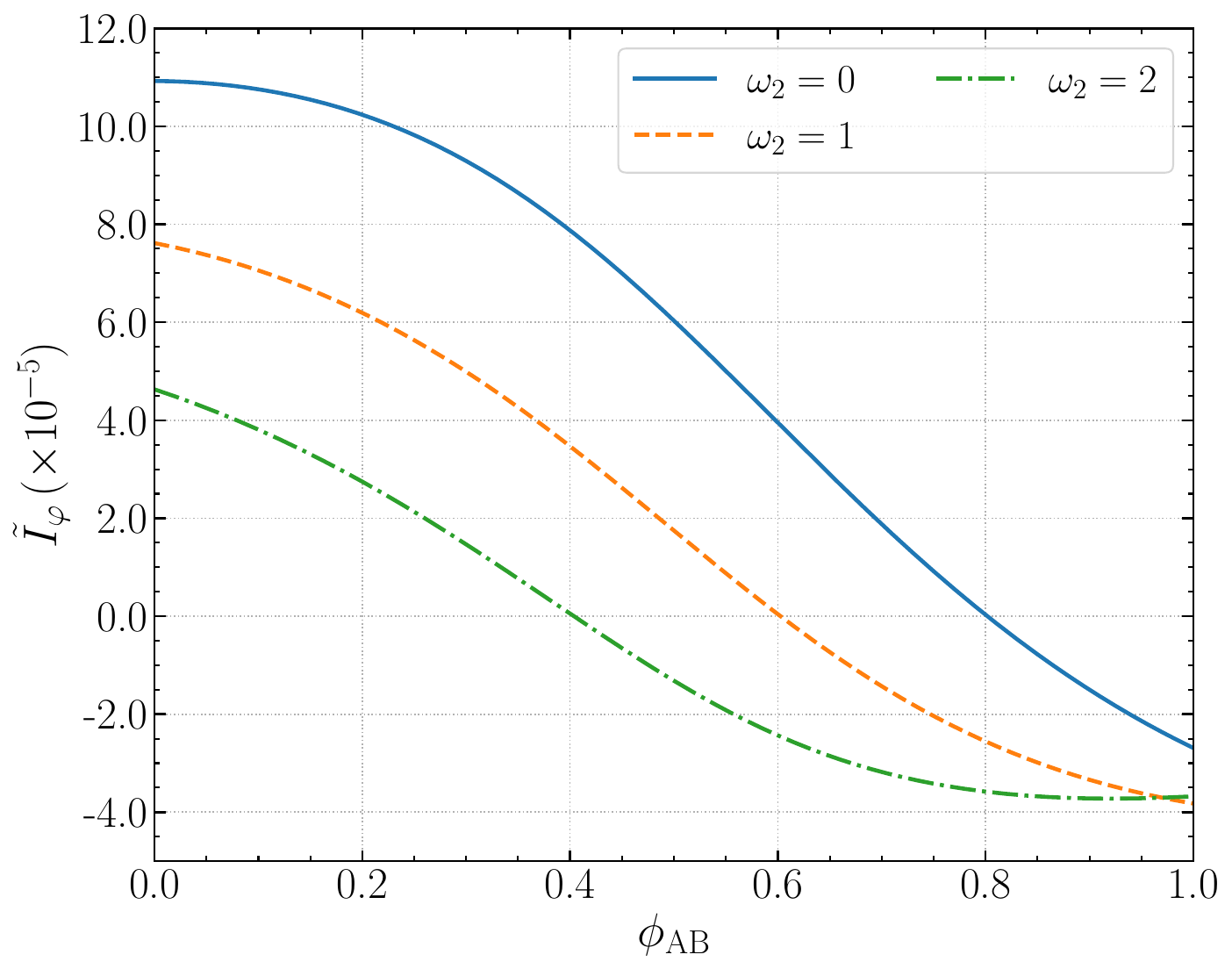}
		\caption{Annularly integrated reduced azimuthal current
			$\tilde I_{\varphi}$ as a function of the reduced AB flux $\phi_{\rm AB}=\Phi/\Phi_{0}$ for three values of the local twist $\omega_{2}\in\{0,1,2\}$ (legend). Parameters are those used in the current section unless otherwise stated: $m^{\ast}/m_{e}=0.067$, $\ell=1$, $k_{z}=0.01~\mathrm{nm}^{-1}$, $\omega_{1}=50~\mathrm{nm}$, and $B=1~\mathrm{T}$ (electron). The curves exhibit the expected AB periodicity $\Delta\phi_{\rm AB}=1$; increasing $\omega_{2}$ produces a mild amplitude renormalization, consistent with the $r$-dependent mixing induced by the twist.}
		\label{fig:Iphi_vs_phiAB}
	\end{figure}
	
	In the numerics we proceed consistently with the spectral solver used for the energy levels (Sec.~\ref{sec:num_methods_results}): we compute $u_{n}(r)$ from the Langer form of the radial Sturm--Liouville problem in Eq.~(\ref{eq:Sl_U_en}), normalize $\int|u_{n}|^{2}\,dr=1$, and then insert $|u_{n}|^{2}/r$ into Eq.~(\ref{eq:currents_components}). This guarantees that current profiles and integrated signals are derived from the same eigenstates and that the discrete version of the continuity law (\ref{eq:cont-law}) remains satisfied on the numerical grid. The qualitative trends seen in the figures follow directly from these formulas: increasing $\omega_{2}$ tilts and redistributes the radial weight (lowering the characteristic radius of $j^{\varphi}$ and shifting the maximum of $j^{z}$), increasing $|B|$ tightens the Landau-like confinement (enhancing the peak magnitude while narrowing its support), and sweeping $\phi$ or $\omega_{1}$ produces persistent-current oscillations consistent with the underlying reindexing symmetry.

	\section{Conclusions}\label{sec:conclusions}
	
	We have presented a unified, gauge-covariant treatment of non-relativistic quantum dynamics in a hybrid screw-twist spacetime with two independent parameters $(\omega_{1},\omega_{2})$. The spatial metric induces a geometric minimal-coupling rule, $\partial_{\varphi}\!\to\!\partial_{\varphi}-f(r)\partial_{z}$ with $f(r)=\omega_{1}+\omega_{2}r$, which cleanly separates global (AB-like) and local (curvature-induced) effects. Combining this geometry with external electromagnetic fields gives a compact radial eigenvalue problem where $\omega_{1}$ shifts the effective angular index, $\omega_{2}$ mixes radial and azimuthal motion through $1/r$ and linear-in-$r$ terms, and $B$ adds Landau-like confinement. 
	
	We derived the continuity equation on the spatial manifold and obtained closed expressions for the azimuthal and axial probability currents, which we used to interpret numerical trends and to clarify the occurrence of near-axis backflow. In the numerics, the axis is regularized by a small core excision with Dirichlet at $r_{\min}$; results are robust under variations of $r_{\min}$ and mesh refinement, showing negligible changes in the lowest eigenvalues and current profiles. Dimensionless control parameters were identified to facilitate reproducibility, and robust numerical strategies (Langer transform, sparse solvers) were deployed. The resulting benchmarks, AB periodicity, Landau fans, and pure screw/twist limits, provide reference points for future work.
	
	The hybrid screw-twist textures discussed here are naturally realized in twisted photonic/electronic waveguides, rolled-up nanomembranes, and metamaterials. Cold-atom platforms offer additional flexibility via synthetic gauge fields and tunable $k_{z}$. Our formulas map the parameters $(\omega_{1},\omega_{2},B,\Phi)$ onto measurable shifts in spectra, interference phases, and persistent currents, and can be readily adapted to include spin, disorder, interactions, or time-periodic drives. On the mathematical side, promising directions include a systematic study of boundary regularization in bounded domains with $\omega_{2}\neq 0$ and the development of semiclassical quantization schemes in the presence of the linear-in-$r$ geometric term.
	
	\section*{Acknowledgments}
	
	F.A. acknowledges the Inter University Centre for Astronomy and Astrophysics (IUCAA), Pune, India for granting visiting associateship. E. O. Silva acknowledges the support from grants CNPq/306308/2022-3, FAPEMA/UNIVERSAL-06395/22, FAPEMA/APP-12256/22, and (CAPES) - Brazil (Code 001).
	
	\section*{Data Availability}
	
	No data were generated or created in this article.
	
	\section*{Conflicts of interest statement}
	
	The authors declare no conflicts of interest.
	
	\section*{Author contributions}
	
	Both authors contributed equally to this work, including conceptualization, methodology, software and figures, and writing (original draft, review, and editing).
	
	\section*{References}
	\bibliographystyle{apsrev4-2}
%

\end{document}